\pdfoutput=1

\documentclass[11pt]{article}

\usepackage[]{acl}

\usepackage{times}
\usepackage{latexsym}

\usepackage[T1]{fontenc}

\usepackage[utf8]{inputenc}

\usepackage{microtype}

\usepackage{graphicx}
\graphicspath{ {./images/} }

\usepackage{listings}
\usepackage{caption}
\usepackage{newfloat}
\DeclareFloatingEnvironment[
  fileext=lol,
  name=Line 51 controls this
]{listing}
\usepackage{subcaption}
\DeclareCaptionSubType{listing}

\definecolor{codegreen}{rgb}{0,0.6,0}
\definecolor{codegray}{rgb}{0.5,0.5,0.5}
\definecolor{codepurple}{rgb}{0.58,0,0.82}
\definecolor{backcolour}{rgb}{0.99,0.99,0.96}
\definecolor{keywords}{RGB}{155,0,90}

\lstdefinestyle{mystyle}{
    backgroundcolor=\color{backcolour},   
    commentstyle=\color{codegreen},
    keywordstyle=\color{keywords},
    numberstyle=\tiny\color{codegray},
    stringstyle=\color{codepurple},
    basicstyle=\ttfamily\scriptsize,
    breakatwhitespace=false,         
    breaklines=true,                 
    captionpos=b,                    
    keepspaces=true,                 
    numbers=left,                    
    numbersep=5pt,                  
    showspaces=false,                
    showstringspaces=false,
    showtabs=false,                  
    tabsize=2,
}

\definecolor{backgroundgreen}{rgb}{0.31, 0.78, 0.47}
\definecolor{backgroundorange}{rgb}{.93, 0.57, 0.13}
\usepackage{url}
\usepackage{xspace}
\usepackage{multirow}
\usepackage{hyperref}
\usepackage{ulem}
\usepackage{amsmath}
\usepackage{subcaption}
\usepackage{booktabs}
\usepackage{float}
\newcommand{\eg}{\hbox{\textit{e.g.,}}\xspace}
\newcommand{\ie}{\hbox{\textit{i.e.,}}\xspace}

\newcommand{\stack}{\textsc{Stack}\xspace}
\newcommand{\pile}{\textsc{Pile}\xspace}

\usepackage{array}
\newcommand{\PreserveBackslash}[1]{\let\temp=\\#1\let\\=\temp}
\newcolumntype{C}[1]{>{\PreserveBackslash\centering}p{#1}}
\newcolumntype{R}[1]{>{\PreserveBackslash\raggedleft}p{#1}}
\newcolumntype{L}[1]{>{\PreserveBackslash\raggedright}p{#1}}

\title{Quantifying Contamination in Evaluating Code Generation \\ Capabilities of Language Models}
\author{Martin Riddell \qquad Ansong Ni \qquad Arman Cohan \\
        Department of Computer Science, Yale University \\ 
        \texttt{\{martin.riddell, ansong.ni,  arman.cohan\}@yale.edu}}

\begin{document}
\maketitle
\begin{abstract}
While large language models have achieved remarkable performance on various code generation benchmarks,
there have been growing concerns regarding potential contamination of these benchmarks as they may be leaked into pretraining and finetuning data.
While recent work has investigated contamination in natural language generation and understanding
tasks, there has been less extensive research into how data contamination impacts the evaluation of
code generation, which is critical for understanding the robustness and reliability of LLMs in programming contexts.
In this work, we perform a comprehensive study of data contamination of popular code generation benchmarks, and precisely quantify their overlap with pretraining corpus through both surface-level and semantic-level matching.
In our experiments, we show that there are substantial overlap between popular code generation benchmarks and open training corpus, and models perform significantly better on the subset of the benchmarks where similar solutions are seen during training.
We also conduct extensive analysis on the factors that affects model memorization and generalization, such as model size, problem difficulty, and question length.
We release all resulting files from our matching pipeline for future research\footnote{Code and data available at \url{https://github.com/yale-nlp/code-llm-contamination}}.
\end{abstract}

\section{Introduction}

\begin{figure}[ht]
  \centering
  \begin{subfigure}[b]{.49\textwidth}
    \centering
    \captionsetup{width=.9\textwidth}
    \includegraphics[width=\textwidth]{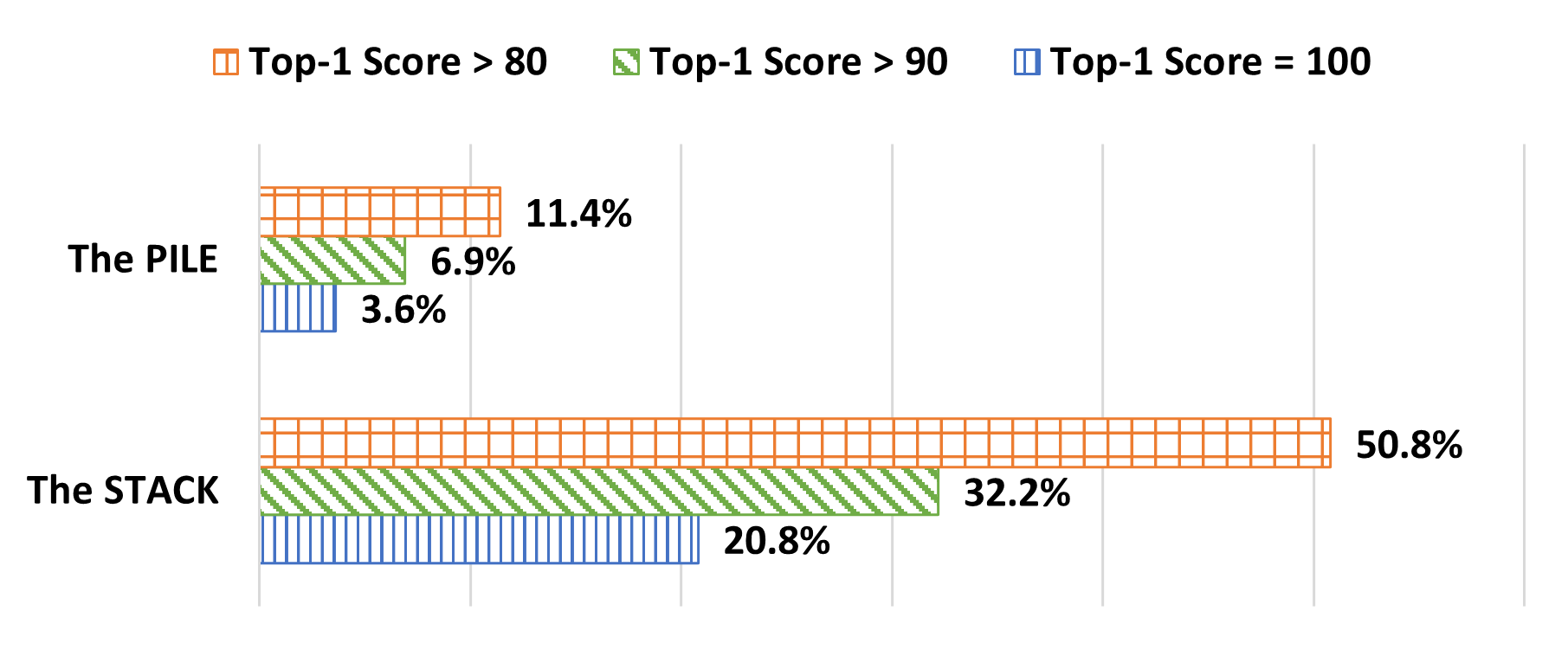}
    \caption{Data contamination on the MBPP benchmark.}
    \vspace{10pt}
    \label{fig:sub1}
  \end{subfigure}
  \begin{subfigure}[b]{.49\textwidth}
    \centering
    \captionsetup{width=.9\textwidth}
    \includegraphics[width=\textwidth]{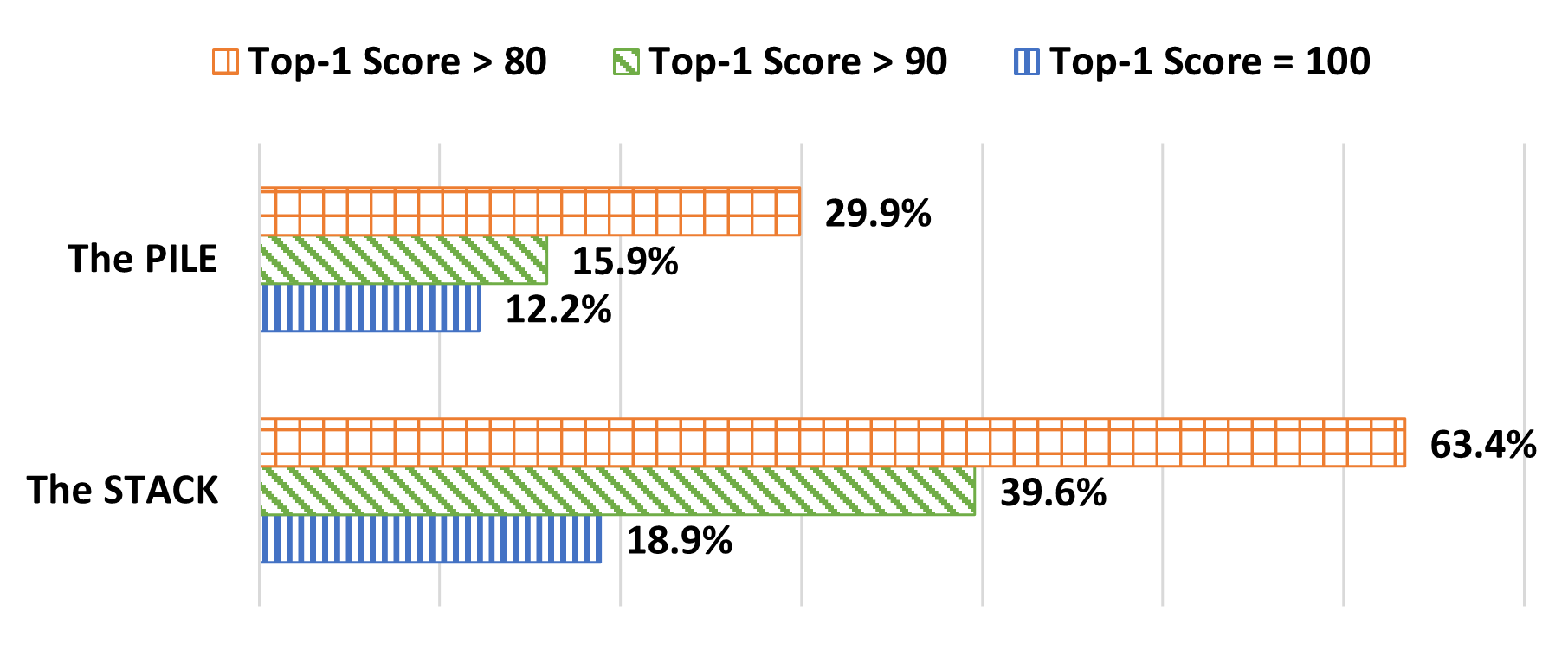}
    \caption{Data contamination on the HumanEval benchmark.}
    \label{fig:sub2}
  \end{subfigure}
  \caption{Quantifying data contamination for the \pile and the \stack corpus on two popular benchmarks, MBPP and HumanEval. ``Top-1 Score'' denotes the  similarity score between the gold solution and the most similar program found in the training corpus.
  }
  \label{fig:cover-fig}
\end{figure}

The compute requirements (encompassing both model size and data volume) for training large language models (LLMs) has grown significantly over the years, correlating with consistent observed enhancements in model performance in both language \cite{kaplan2020scaling,hoffmann2022training} and code \cite{ni2023l2ceval} generation tasks.
Larger models trained on larger training corpora tend to lead to an increased risk of \textit{data contamination}, which we 
refer to as instances of evaluation benchmark data appearing within the data used during the training of models. 
LLMs tend to perform better on evaluation samples  that resemble the documents and instances encountered during training \cite{Kandpal2022LargeLM, razeghi-etal-2022-impact, magar-schwartz-2022-data}, and are more likely to emit memorized training data when they have seen it multiple times \cite{kandpal2022deduplicating, carlini2023quantifying}. Recent papers have also shown evidence that LLMs are possibly contaminated \cite{golchin2023time, yang2023rethinking}, which limits our understanding of their generalization capabilities to unseen scenarios.

Despite significant research into data contamination in natural language (NL) benchmarks \cite{golchin2023time, chang2023speak, blevins-zettlemoyer-2022-language, dodge-etal-2021-documenting}, there's been relatively little exploration into how this issue affects the evaluation of \textit{code generation} capabilities in LLMs. We posit that the fundamental disparities between NL and programs warrant a deeper examination. Recent studies, such as work by \citet{karmakar2022codex, ranaldi2024investigating}, suggest that code-based LLMs may demonstrate patterns of memorization, underscoring the need for scrutiny into their generalization capabilities to unseen cases. Key distinctions between code and NL include the critical role of syntax and the variable requirements for naming functions and variables across different programs. These differences lead us to argue that traditional surface-level comparisons might not be adequate for identifying contaminated data points.

In this paper, we propose a pipeline to measure the overlap between code generation benchmarks and pretraining corpus of code LLMs, incorporating both surface-level and semantic-level program matching. 
As a result of the exhausive search amount the training corpus with our pipeline, we provide a precise quantification of the examples whose solutions are seen during training, for popular code generation benchmarks as MBPP \cite{austin2021program} and HumanEval \cite{chen2021evaluating}.
We study two popular open pretraining corpus which contain code, the \pile \cite{gao2020pile} and the \stack, as well as three model series trained on either corpora, StarCoderBase \cite{li2023starcoder}, Pythia \cite{pmlr-v202-biderman23a} and CodeGen-NL \cite{nijkamp2023CodeGen}.
Our results demonstrate
severe contamination of the widely used MBPP and HumanEval benchmarks within the \pile and the \stack corpora, as shown in \autoref{fig:cover-fig}, with models performing significantly better on questions that the models have seen the same or similar program solutions to.
We also perform thorough analysis on factors that may affect model memorization and generalization such as model size and difficulty of the questions.
We also include a case study on outliers, to provide a more comprehensive understanding of model behavior given different levels of exposure to test data.

\section{Methodology}
To quantify data contamination for code LLMs, we first introduce methods used to measure program similarity from surface- and semantic-level in \autoref{sec:measuring-similarity}. Next, in \autoref{sec:quantifying-contamination} we describe how to combine similarity measurements to identify the overlapping programs in the training data and test benchmarks as well as introduce how to quantify data contamination based on the similarity scores and the number of appearances of similar programs seen during the course of training. %

\subsection{Measuring Program Similarity}
\label{sec:measuring-similarity}

While most popular code generation benchmarks focus on generating functions, the training data are often chunked by files, which may contain multiple functions or classes. This means that
document-level de-duplication techniques \citep[e.g.,][]{10.1145/3359591.3359735} cannot be used effectively, as other programs within the document may add too much noise. Thus we opt to perform substring-level matching, which is much more computationally heavy but also more accurate than methods used by previous work \cite{Lee2022DoLM, Peng2023NearDuplicateSS, Kandpal2022LargeLM}.
More specifically, we use a sliding window to scan the training data character-by-character and compute its similarity scores with gold solutions in the benchmarks.
To maximize the recall of possible contaminated examples in coding benchmarks,
we employ both surface- and semantic-level similarity measurements. 

\paragraph{Surface-Level Similarity.}
\label{sec:surface-similarity}
To measure surface-level similarity between programs, we use the \textit{Levenshtein similarity score} \cite{sarkar-etal-2016-junitmz},
which is the Levenshtein edit distance \cite{Levenshtein1965BinaryCC} normalized by the length of both the source and target strings.
We selected the Levenshtein similarity score as the first step in our pipeline because it is an easy-to-compute and intuitive measurement that can handle surface-level fuzzy matches between two programs. While the Levenshtein edit distance has been used before to deduplicate datasets at a file level \cite{chowdhery2022palm}, we perform it on a substring level. 
An example of this can be found in \autoref{fig:levenshtein_program_examples}.\footnote{we use the \texttt{rapidfuzz} python library to calculate the similarity score \url{https://pypi.org/project/rapidfuzz/}.}

\paragraph{Semantic Similarity.}
\label{sec:semantic-similarity}
 While the surface-level similarity metrics can easily capture similar programs in surface form, two semantically similar or even identical programs can have very different surface form due to different identifiers (\eg variable names) or whitespace characters. 
Therefore, finding semantically similar programs is also crucial for measuring contamination and understanding the generalization capabilities of the models. 
To measure semantic similarity between programs, we adopt the \textit{Dolos toolkit} \cite{maertens2022dolos}, which is a source code plagiarism detection tool for education purposes. Dolos first uses \texttt{tree-sitter}\footnote{\url{https://tree-sitter.github.io/tree-sitter/}} to tokenize and canonicalize the program into representations of abstract syntax trees (ASTs), then computes a similarity score representing the semantic-level similarity based on the $k$-gram matching between source and target programs. Since Dolos measures similarities based on the ASTs, non-semantic changes that greatly decrease the Levenshtein similarity scores, such as indentations and variable/function names, will not affect the scores calculated by Dolos. An example of this can be found in \autoref{fig:dolos_program_example}.
Dolos was also used in previous works for detecting intellectual property violations \cite{pmlr-v202-yu23g}.

\begin{figure}[!t]
    \centering
    \include{tabs/levenshtein_program_examples}
    \vspace*{-5mm}\caption{Example where surface-level matching works better than semantic-level. 
    Because most of the program is commented out, the semantic-level similarity score is 0 despite the programs being otherwise identical.
    }
    \label{fig:levenshtein_program_examples}
\end{figure}

\begin{figure}[!t]
  \centering
  \begin{subfigure}[b]{.22\textwidth}
    \centering
    \includegraphics[scale=.45]{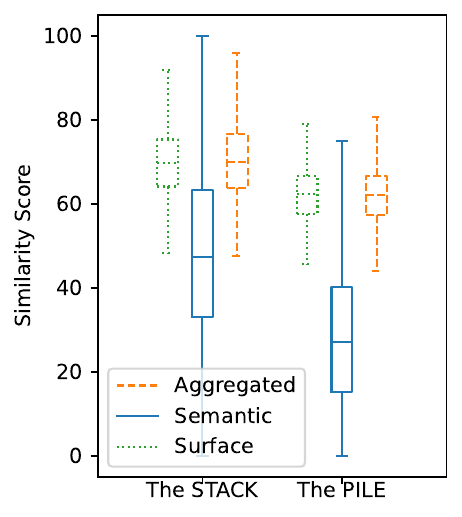}
    \caption{Top-10 Scores.}
    \label{fig:sub1}
  \end{subfigure}
  \begin{subfigure}[b]{.22\textwidth}
    \centering
    \includegraphics[scale=.45]{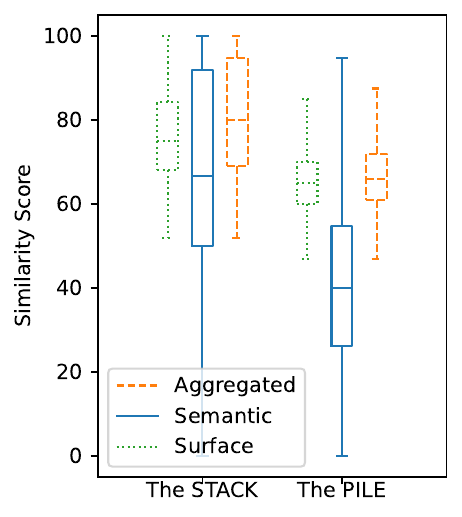}
    \caption{Top-1 Score.}
    \label{fig:sub2}
  \end{subfigure}
  \caption{Distribution of different similarity scoring methods on the MBPP dataset. Similar results for HumanEval are shown in \autoref{fig:HE_relevant_score_info}.
}
  \label{fig:relevant_score_info}
\end{figure}

\subsection{Quantifying Data Contamination}
\label{sec:quantifying-contamination}

For each problem and its gold solution in the test benchmark (\eg MBPP), we would like to determine the most similar programs that the models have seen during training. 
However, this would require us to perform a pair-wise comparison with all programs in the training data using the similarity score metrics mentioned in \S\ref{sec:measuring-similarity}. Because training data is usually on the scale of hundreds of gigabytes to terabytes, it is computationally expensive\footnote{
We estimate that it will take $5.2*10^5$ CPU hours to search just the Python files from the \stack for MBPP.
} to run surface-level matching methods; running the code-specific semantic matching methods on the entire training dataset is even more computationally prohibitive.
\paragraph{Aggregating Similarity Scores.}
We use a two-stage process to analyze test examples and their correct (gold standard) program solutions. First, we measure the surface-level similarity by calculating Levenshtein scores. This involves comparing all substrings of the same length as the gold solution across all relevant files in specific subsets of our dataset (see \autoref{sec:exp-setup-models} for details).
We keep the top 500 programs\footnote{This is determined by a combination of automatic and manual inspection. For example, at the $500^{th}$ most similar program from the \stack for MBPP, 95\% of them have a similarity score $<72$, which is no longer relevant by human inspection.} with the highest Levenshtein similarity scores for each test example for the next step.
The similarity scores found by searching the \pile and the The \textsc{Stack} for gold programs from the MBPP benchmark are shown in \autoref{fig:relevant_score_info}, along with a comparison between results found using only the top score and those found using an average of the top 10 scores.
With the top 500 programs 
with the highest Levenshtein similarity scores from the training data, we further compute the semantic similarity scores with the gold programs using Dolos. Then the \textit{aggregated similarity score} is computed as the maximum of the surface-level similarity score ($S_{\mathtt{surface}}$) and semantic similarity score ($S_{\mathtt{semantic}}$) similarity scores:
\begin{equation*}
    S(p, p^*) = \max(S_{\mathtt{surface}}(p, p^*), S_{\mathtt{semantic}}(p, p^*))
\end{equation*}
This aggregated similarity score is a simple and intuitive way to reflect how programs can be similar both from their surface form and semantics.

\begin{figure}[!t]
    \centering
    \include{tabs/dolos_program_examples}
    \vspace*{-5mm}\caption{Example where semantic-level similarity works better than surface-level. The ASTs of two programs are identify despite different variable names. 
    }
    \label{fig:dolos_program_example}
\end{figure}

\section{Experimental Setup}
We select two of the most popular public pretraining corpora for general LLM and code LLMs, namely the \pile \cite{gao2020pile} and the \stack \cite{kocetkov2022stack}, and three series of popular open-source models, \ie Pythia \cite{pmlr-v202-biderman23a}, CodeGen-NL \cite{nijkamp2023CodeGen} and StarCoderBase\footnote{%
StarcoderBase refers to the models originally trained on the \stack. These models were further finetuned on Python code to create the Starcoder models \cite{li2023starcoder}.}\cite{li2023starcoder}. 
For the coding benchmark, we opt to study MBPP \cite{austin2021program} and HumanEval \cite{chen2021evaluating} due to their popularity.
We introduce them in more detail in the following subsection.

\begin{table*}[]
\small
\setlength{\tabcolsep}{4pt}
\begin{subtable}[t]{\textwidth}
\centering

\begin{tabular}{llrrrrrrrrr}
\toprule
\multicolumn{1}{c}{\multirow{2}{*}{\textbf{Benchmark}}} & \multicolumn{1}{c}{\multirow{2}{*}{\textbf{Models}}} & \multicolumn{1}{c}{\textbf{}} & \multicolumn{2}{c}{\textbf{Top-1 Score=100}}           & \multicolumn{1}{c}{\textbf{}} & \multicolumn{2}{c}{\textbf{Top-1 Score\textgreater{}90}} & \multicolumn{1}{c}{\textbf{}} & \multicolumn{2}{c}{\textbf{Top-1 Score\textgreater{}80}} \\ \cline{4-5} \cline{7-8} \cline{10-11} 
\multicolumn{2}{c}{}                                 & \multicolumn{1}{c}{$Acc_o$}     & \multicolumn{1}{l}{\% Rm} & \multicolumn{1}{c}{$Acc_d$} & \multicolumn{1}{l}{}          & \multicolumn{1}{l}{\% Rm}  & \multicolumn{1}{c}{$Acc_d$}  & \multicolumn{1}{l}{}          & \multicolumn{1}{l}{\% Rm}  & \multicolumn{1}{c}{$Acc_d$}  \\ \midrule
\multirow{3}{*}{\textbf{MBPP}} & StarCoderBase-15.5B                                  & 41.6                        & 20.8                      & 33.8 (-18.8\%)             &                               & 32.2                       & 32.5 (-22.6\%)                &                               & 50.8                       & 29.7 (-28.6\%)               \\
& Pythia-12B                                           & 17.8                        & 3.6                       & 17.0 \phantom{1}(-4.5\%)               &                               & 6.9                        & 16.6 \phantom{1}(-6.7\%)                &                               & 11.4                       & 15.8 (-11.2\%)                 \\
& CodeGen-NL-16B                                       & 19.6                        & 3.6                       & 18.4 \phantom{1}(-6.1\%)             &                               & 6.9                        & 17.4 (-11.2\%)                &                               & 11.4                       & 16.5 (-15.8\%)                \\ \midrule
\multirow{3}{*}{\textbf{HumanEval}} & StarCoderBase-15.5B                                  & 30.5                          & 18.9                      & 22.6 (-25.9\%)                &                               & 39.6                       & 15.2 (-50.2\%)                &                               & 63.4                       & 20.0 (-34.4\%)                \\
& Pythia-12B                                           & 9.8                           & 12.2                      & 4.2 (-57.1\%)                &                               & 15.9                       & 2.9 (-70.4\%)                 &                               & 29.9                       & 1.7 (-82.7\%)                  \\
& CodeGen-NL-16B                                       & 14.6                          & 12.2                      & 8.3 (-43.2\%)                &                               & 15.9                       & 5.8 (-60.3\%)                 &                               & 29.9                       & 3.5 (-76.0\%)                \\ \bottomrule
\end{tabular}

\label{tab:table1_c}
\end{subtable}
\caption{Measuring the de-contaminated accuracy ($Acc_d$) by removing potentially contaminated subsets of MBPP and HumanEval \textit{w.r.t.} different thresholds. ``$Acc_o$'' denotes original model accuracy and ``\% Rm'' denotes the percentage of the dataset removed. The \textit{relative} accuracy degradation after de-contamination is shown in brackets. 
}
\label{tab:decontamination-results-both}
\end{table*}

\subsection{Models and Pretraining Data}
\label{sec:exp-setup-models}
We select the models by the following criteria:
\textit{1)} The pretraining data for the models must be publicly available;
\textit{2)} To ensure non-trivial performance on the coding benchmarks, such models must have Python code in their pretraining data;
\textit{3)} Additionally, we do not consider any models that are instruction-tuned, or trained with reinforcement learning from human feedback (\ie RLHF), as it is hard to quantify the effect of such instruction-tuning/human preference data along with the pretraining corpus. Based on these criteria, we study the following three model series in this work:

\textbf{The \textsc{Pile} and Pythia.} \hspace{0.7em} 
Pythia~\citep{pmlr-v202-biderman23a} is a suite of 16 LLMs intended to facilitate research in many areas. All models are trained on the \pile dataset \cite{gao2020pile}, with their size ranging from 70M to 12B parameters. We used the 1.4B, 2.8B, 6.9B, and 12B models for this study. We use the GitHub split of the training dataset, which has a raw size of 95.16 GiB.

\textbf{The \textsc{Pile} and CodeGen-NL.} \hspace{0.7em} 
Another series of models that are trained with \pile is CodeGen-NL \cite{nijkamp2023CodeGen}, and we study the 350M, 2B, 6B, and 16B versions of it.
Though stronger CodeGen models are available via further training on more code data, the exact copy of such data is not publicly released thus we choose to study the CodeGen-NL series.
Due to the overlap of training data, we use the results of searching through the GitHub split that we did for the Pythia models.

\textbf{The \textsc{Stack} and StarCoderBase.} \hspace{0.7em} 
We use the 1B, 3B, 7B and 15.5B StarCoderBase models \cite{li2023starcoder} that were trained on the \textsc{Stack} dataset \cite{kocetkov2022stack}. Due to the size of the training data, we only search through 60.40 GB within the Python split of its training dataset.
The \stack was created from permissively-licensed source code files, and was open-sourced to make the training of code LLMs more reproducible.\footnote{It is worth noticing that \stack went through a string-matching-based decontamination process for MBPP and HumanEval, but we are still able to find traces of contamination for these two datasets. }

\subsection{Benchmarks}
We measure the data contamination issues for the following two popular coding benchmarks:

\textbf{MBPP} \cite{austin2021program} is a benchmark containing 974 short, crowd-sourced Python programming problems. We use the 500 questions within its test split. 

\textbf{HumanEval} \cite{chen2021evaluating} is a benchmark consisting of 164 hand-written problems. Each problem contains a gold solution.

Notably, these two benchmarks come with gold program solutions, which we use to search the pretraining data as a query.
To obtain the model performance and predictions on each of the dataset examples, we use the evaluation framework and model outputs from \texttt{L2CEval} \cite{ni2023l2ceval}.

\begin{table}[]
\small
\setlength{\tabcolsep}{3pt}
\centering
\begin{tabular}{lrrrrrr}
\toprule
\multicolumn{1}{c}{\multirow{2}{*}{\textbf{Models}}} &  \multicolumn{3}{c}{\textbf{MBPP}}                                                      & \multicolumn{3}{c}{\textbf{HumanEval}}                                                 \\ 
 & \multicolumn{1}{c}{$\Uparrow^{10\%}$} & \multicolumn{1}{c}{$\Downarrow_{10\%}$} & \multicolumn{1}{c}{$\Delta_\Updownarrow$} & \multicolumn{1}{c}{$\Uparrow^{10\%}$} & \multicolumn{1}{c}{$\Downarrow_{10\%}$} & \multicolumn{1}{c}{$\Delta_\Updownarrow$}\\ \midrule
 StarCoderBase & 72.0 & 22.0 & 50.0 & 75.0 & 31.3 & 43.7 \\
Pythia        & 40.0 & 8.0  & 42.0 & 56.3 & 0.0  & 56.3 \\
CodeGen-NL    & 48.0 & 6.0  & 42.0 & 62.5 & 0.0  & 62.5 \\\bottomrule
\end{tabular}
\caption{We show the performance gap ($\Delta_\Updownarrow$) between the top 10\% ($\Uparrow^{10\%}$) and bottom 10\% ($\Downarrow_{10\%}$) of programs based on the average of the top-10 aggregated similarity scores. %
Only the \textit{largest} models are shown for each model series, full results available in \autoref{tab:performance_gap_full}.
}

\label{tab:performance_gap_both}
\end{table}
\begin{figure*}[htb]
    \centering %
\begin{subfigure}{0.33\textwidth}
  \includegraphics[width=1.\linewidth]{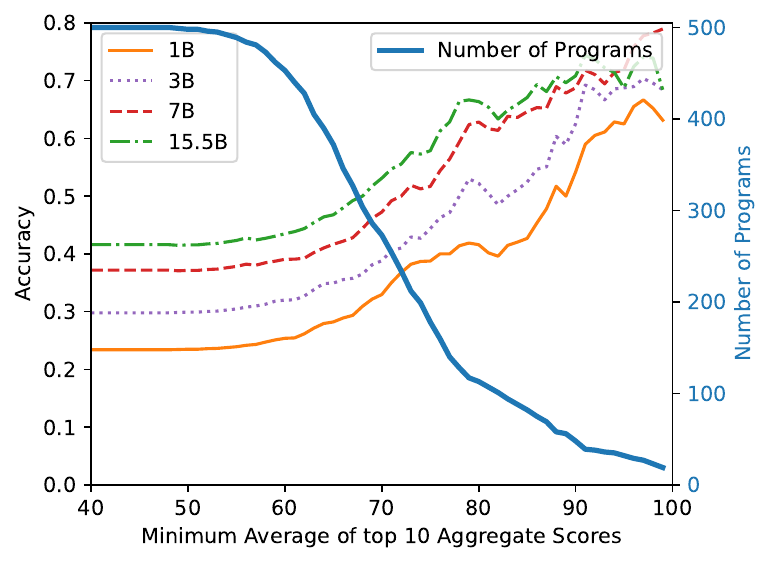}
  \caption{StarCoderBase on MBPP}
  \label{fig:1}
\end{subfigure}\hfill %
\begin{subfigure}{0.33\textwidth}
  \includegraphics[width=1.\linewidth]{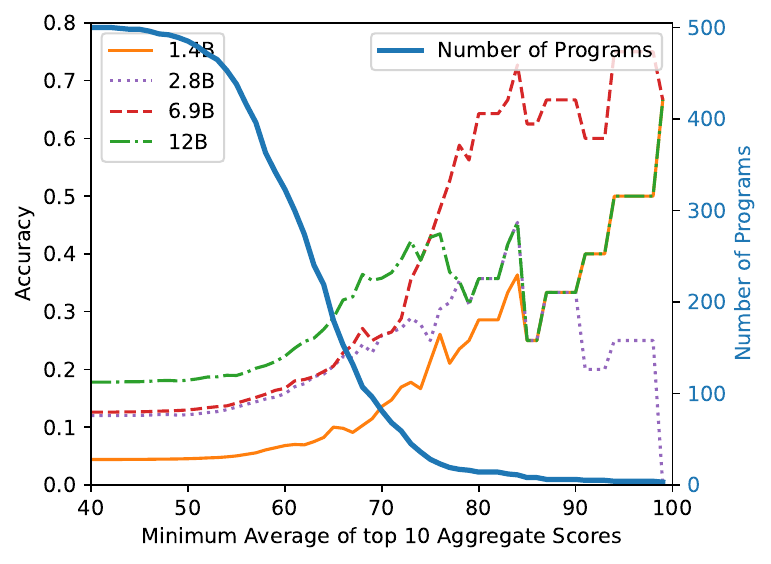}
  \caption{Pythia on MBPP}
  \label{fig:2}
\end{subfigure}\hfill %
\begin{subfigure}{0.33\textwidth}
  \includegraphics[width=1.\linewidth]{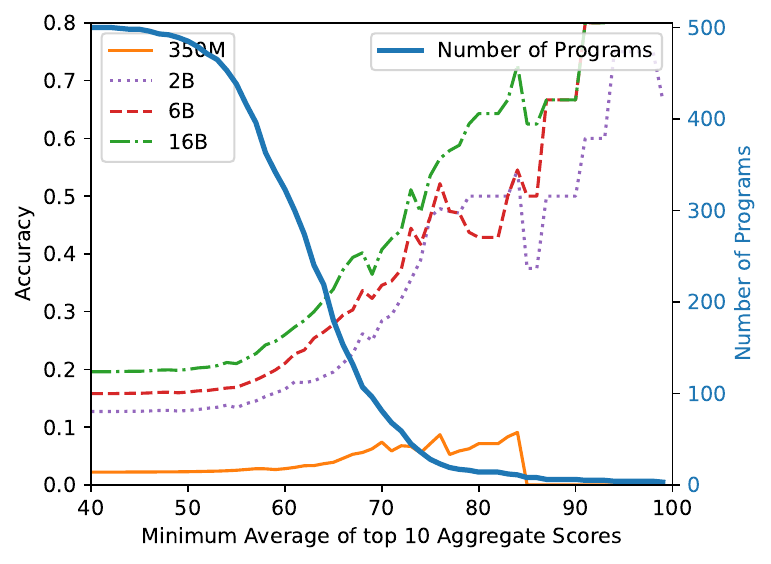}
  \caption{CodeGen-NL on MBPP}
  \label{fig:3}
\end{subfigure}

\medskip
\begin{subfigure}{0.33\textwidth}
  \includegraphics[width=1.\linewidth]{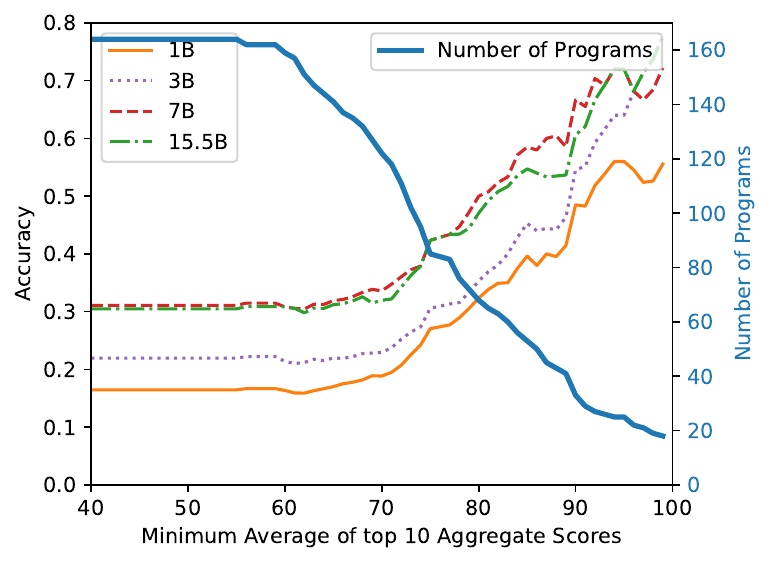}
  \caption{StarCoderBase on HumanEval}
  \label{fig:4}
\end{subfigure}\hfill %
\begin{subfigure}{0.33\textwidth}
  \includegraphics[width=1.\linewidth]{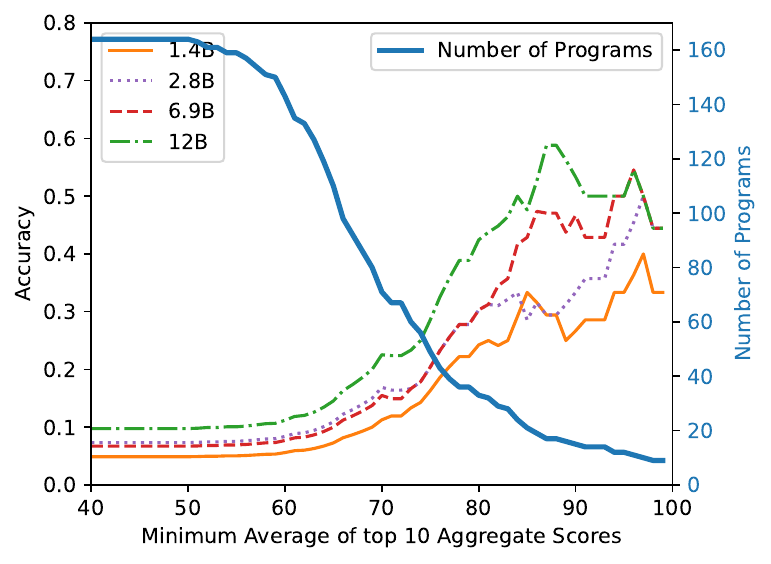}
  \caption{Pythia on HumanEval}
  \label{fig:5}
\end{subfigure}\hfill %
\begin{subfigure}{0.33\textwidth}
  \includegraphics[width=1.\linewidth]{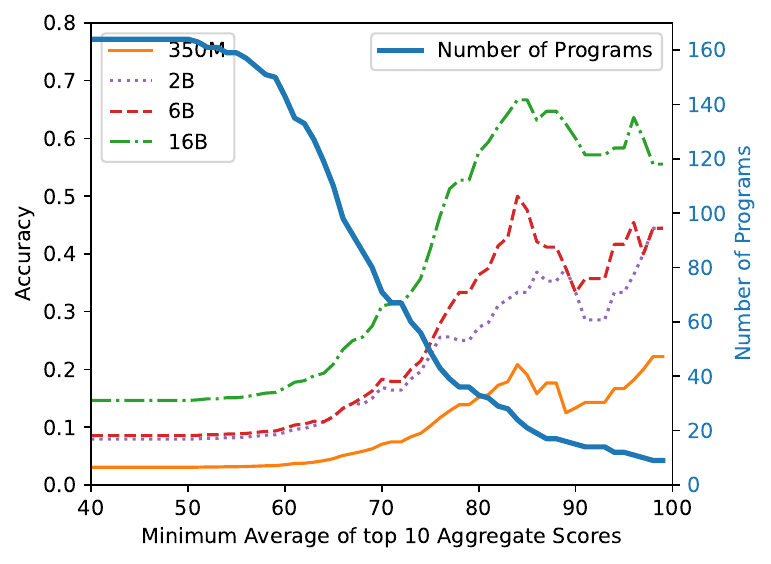}
  \caption{CodeGen-NL on HumanEval}
  \label{fig:6}
\end{subfigure}
\caption{Accuracy of different model series evaluated on a subset of examples with increasing overlap with the model's pretraining data. Subset obtained by using the $x$-axis as a threshold for the minimum score obtained by taking the average aggregated similarity score of top-10 matched programs in the training data.
}
\label{fig:both_aggregate_scores_on_mbpp_full}
\end{figure*}

\section{Results}

In this section, we first present our main results in \S\ref{sec:main-results}, then we perform several analysis on how the length, difficulty and model sizes affects the our findings in \S\ref{sec:analysis}, and finally we present a case study in \S\ref{sec:case-study}.

\subsection{Main Results}
\label{sec:main-results}

\paragraph{3.6\% to 20.8\% of the solutions are seen during training.}
For an example in the test data (\ie those of MBPP or HumanEval), we note it as ``\textit{seen}'' if the aggregated similarity score is 100, \ie a perfect match exists on the surface- or semantic-level. 
Results in \autoref{fig:cover-fig} show that 12.2\% of the solutions in HumanEval have been seen by models trained on the \pile and 18.9\% have been seen by models trained on the \stack. 
For MBPP, 3.6\% of it can be found in the \pile while as much as 20.8\% have been seen by models trained on the \stack.
Much less overlap is found for the \pile, as 3.6\% of MBPP, but 20.8\% of the solutions on MBPP problems have been seen for models trained on the \stack.
These results suggest that a non-trivial part of both MBPP and HumanEval have been seen for the models trained on either the \pile or the \stack, suggesting a high contamination rate.

\paragraph{Models perform significantly better when similar solutions are seen during training.}
To highlight the difference in performance that models have between questions which they have seen similar solutions and questions which they have not, following \citet{razeghi-etal-2022-impact}, we use the performance gap between the top 10\% most similar and bottom 10\% least similar programs.
The performance gap of models from the chosen model series is shown in \autoref{tab:performance_gap_both}, where it can be observed that all three models perform significantly better on questions to which the solution is in the top 10\% of compared to questions where the solution is in the bottom 10\%.
StarCoderBase-15.5B achieves an accuracy of 72\% on the top 10\% of questions and an accuracy of 22\% on the bottom 10\% of questions of the MBPP benchmark.
The range of similarity scores for each model and benchmark can be found in \autoref{fig:relevant_score_info}. In \S\ref{sec:analysis} we provide an analysis on decoupling memorization and question difficulty. 

\paragraph{De-contaminated results.}
In an attempt to show the impact of seen questions on model performance, we remove potentially contaminated questions from each benchmark, showing the results in \autoref{tab:decontamination-results-both}.
We observe that removing not only questions that have been seen, but also questions where programs similar to the gold program have been seen during training has an adverse effect on model performance.
Moreover, from the de-contaminated results, the performance gap between different models could be much smaller. For example, the original accuracy ($Acc_o$) gap between StarCoderBase-15.5B and Pythia-12B is 23.8\%, and after de-contamination, the performance gap is decreased to 13.9\%. This indicates that a large part of the performance gap between different models may due to data contamination. While we do not find the performance rankings of the models to change with de-contaminated results, a study with more models might be needed for deriving any general conclusions.

\begin{table*}[htbp]
    \centering
    \begin{subtable}[b]{0.65\textwidth}\centering
\small
\setlength{\tabcolsep}{3pt}
\centering
\begin{tabular}{l|rrrr|rrrr}
\toprule
\multirow{2}{*}{\textbf{Models}}      & \multicolumn{4}{c|}{\textbf{MBPP}}     &                                                                                \multicolumn{4}{c}{\textbf{HumanEval}}                                                                                \\ \cline{2-5}\cline{6-9}
\textbf{} & \multicolumn{1}{c}{\textbf{$\mathcal{D}_S$}} & \multicolumn{1}{c}{\textbf{$\mathcal{D}_{S^-}$}} & \multicolumn{1}{c}{\textbf{$\mathcal{D}_P$}} & \multicolumn{1}{c|}{\textbf{$\mathcal{D}_{P^-}$}} & \multicolumn{1}{c}{\textbf{$\mathcal{D}_S$}} & \multicolumn{1}{c}{\textbf{$\mathcal{D}_{S^-}$}} & \multicolumn{1}{c}{\textbf{$\mathcal{D}_P$}} & \multicolumn{1}{c}{\textbf{$\mathcal{D}_{P^-}$}}\\ \midrule
StarCoderBase-7B    & \textbf{63.5} & 30.3          & 33.3          & \textbf{37.3} & \textbf{64.5} & 23.3 & \textbf{75.0} & 25.0 \\
StarCoderBase-15.5B & \textbf{71.2} & 33.8          & \textbf{55.6} & 41.1          & \textbf{64.5} & 22.6 & \textbf{80.0} & 23.6 \\\midrule
CodeGen-NL-6B       & 11.5          & \textbf{16.9} & \textbf{38.9} & 14.9          & \textbf{29.0} & 3.8  & \textbf{45.0} & 3.5  \\
CodeGen-NL-16B      & 11.5          & \textbf{21.7} & \textbf{50.0} & 18.5          & \textbf{48.4} & 6.8  & \textbf{60.0} & 8.3  \\ \hline
\end{tabular}
\caption{Model performance on different subsets of MBPP and HumanEval.}
\label{tab:mem_decouple}
\end{subtable}
\hfill
\begin{subtable}[b]{0.33\textwidth}\centering
\small
\setlength{\tabcolsep}{3pt}
\centering
\begin{tabular}{lcccc}
\toprule
\multicolumn{1}{l}{\textbf{}}    &\multicolumn{1}{c}{$|\mathcal{D}|$}    & \multicolumn{1}{c}{$|\mathcal{D_S}|$} & \multicolumn{1}{c}{\textbf{$|\mathcal{D}_{P}|$}} & \multicolumn{1}{c}{\textbf{$|\mathcal{D}_{S \cap P}|$}} \\ \midrule
\textbf{MBPP} & 500 & 104                                                         & 18 & 2                                    \\
\textbf{HumanEval} & 164 & 31                                                                & 20              & 16                   \\ \bottomrule
\end{tabular}
\caption{The number of questions seen by models trained on the \stack ($\mathcal{D_S}$) or the \pile ($\mathcal{D_P}$), and the number of programs in both subsets ($|\mathcal{D}_{S \cap P}|$).}
\label{tab:mem_decouple_sizes}
\end{subtable}
\caption{Decoupling memorization and difficulty. 
$\mathcal{D_S}$ denotes the subset that overlaps with the \stack, and $\mathcal{D_{S^-}}$ denotes the complement set (\ie $\mathcal{D_{S^-}}=\mathcal{D} - \mathcal{D_S}$). We define $\mathcal{D_P}$ and $\mathcal{D_{P^-}}$ similarly for the \pile.
The better performance amount the two disjoint subsets (\ie $\mathcal{D_*}$ and $\mathcal{D_{*^-}}$) are in \textbf{bold}.
}
\label{tab:mem_decouple_all}
\end{table*}

\subsection{Analysis}
\label{sec:analysis}

\paragraph{Ablations on model size.}

We show how model size affects accuracy in regard to the aggregate score in \autoref{fig:both_aggregate_scores_on_mbpp_full}. We observe that larger models tend to perform better than smaller models in their family, indicating that they are not only better at generalization, but also at memorization.
We believe that the Pythia and CodeGen-NL models show similar trends in these graphs due to being trained on the same training data.
We also note that the noise in each graph grows as there are fewer programs being evaluated on, explaining the 0\% accuracy that some models show when evaluated on only questions they have seen 10 or more times during training.

\paragraph{Decoupling memorization and difficulty.}
We attempt to show that the model performance on seen questions is not just outlying questions that are easier than the remaining questions in the benchmarks.
To do so, we compare the overall performance of models on the MBPP and HumanEval benchmarks against their performance on different subsets of questions based on seen and unseen questions in \autoref{tab:mem_decouple}.
We show that while StarCoderBase models perform better on the subset of questions in MBPP that they have seen than on the unseen questions, Pythia and CodeGen-NL models generally perform worse on the same subset of questions.
We also provide the sizes of each subset in \autoref{tab:mem_decouple_sizes}, and note the significant overlap between questions in HumanEval that have been seen by the \stack or the \pile as compared to the MBPP benchmark.
The improved performance of models in the StarCoderBase family on familiar questions in the MBPP benchmark does \textit{not} appear to result from these questions being easier than those they haven't encountered during training.
For example, CodeGen-NL-16B has an overall accuracy of 19.6\% on the MBPP benchmark, but has an accuracy of only 11.5\% on the 104 questions that StarCoderBase has seen. 
This indicates that models having seen a solution to a question during training significantly increases the performance of models on these questions.

\paragraph{Effect of program length on similarity scoring.}

\begin{figure}
    \centering
    \includegraphics[scale=.45]{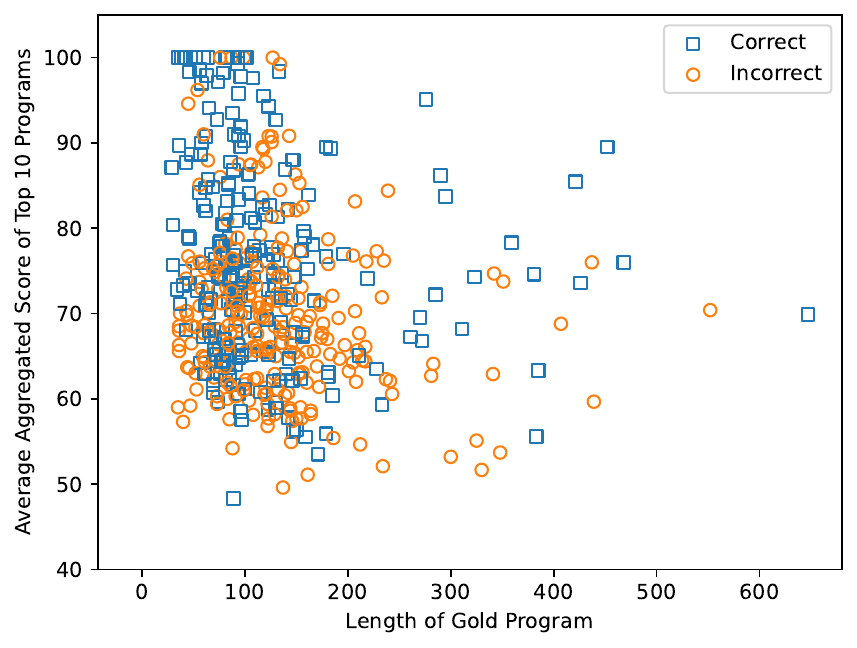}
    \caption{
    Gold solution length \textit{vs.} overlap with training data \textit{vs.} model prediction correctness, for StarCoderBase-15.5B on MBPP. Similar results for HumanEval are shown in \autoref{fig:he_size_against_similarity}.
    }
    \label{fig:size_against_similarity}
\end{figure}

One possible concern is that the size of the gold programs could affect the similarity score. Longer strings can have more differences between one another without affecting their aggregated similarity score as much as in shorter strings. To analyze this, we plot the length of every gold program within the MBPP benchmark against the aggregated similarity score of the most similar string within the training dataset used for the StarCoderBase model family in \autoref{fig:size_against_similarity}. There does not appear to be a correlation between the length and the aggregated similarity score, or length and accuracy.

\subsection{Case Study}
\label{sec:case-study}

\begin{figure}[!t]
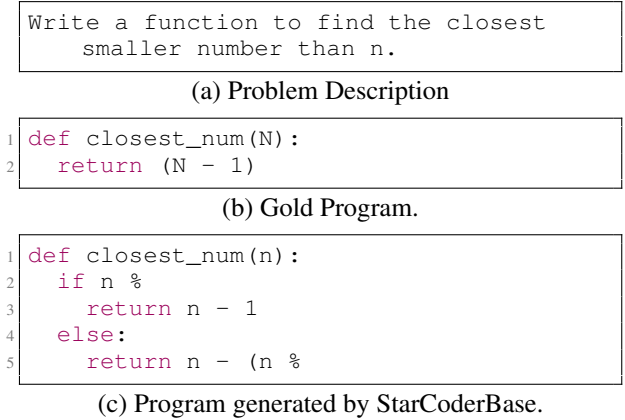

    \centering
    \include{tabs/Case_Study_Unseen_Exemplar}
    \vspace*{-5mm}\caption{Another example where despite similar solutions appearing 10 or more times in the training corpus, StarCoderBase still fails at test time.
    }
    \label{fig:case_study_unseen_exemplar}
\end{figure}

Here we present a case study, by showing examples where the models have seen the gold solutions to 10 or more times but still fails to produce a correct solution at test time. Two representative examples are shown as \autoref{fig:case_study_seen_exemplar} and \autoref{fig:case_study_unseen_exemplar}.
For the first example (\autoref{fig:case_study_seen_exemplar}), although programs that are similar to the gold program appears multiple times in the pretraining data, understanding the problem description is arguably harder part of the problem. As for the second example (\autoref{fig:case_study_unseen_exemplar}), the gold program is quite simple thus it is not surprising that multiple matches in the training corpus are found, but it may also make it difficult for the model to associate such program with any specific natural language description.

\begin{figure}[!t]
    \centering
    \include{tabs/Case_Study_Seen_Exemplar}
    \vspace*{-5mm}\caption{Example where despite similar solutions appearing 10 or more times in the training corpus, StarCoderBase still fails at test time.
    }
    \label{fig:case_study_seen_exemplar}
\end{figure}

\section{Related Work}

\textbf{Measuring contamination.} 
Our study on the effect of contaminated test questions on accuracy are similar to the work done by \citet{Carlini2020ExtractingTD, Henderson2017EthicalCI, jiang2024investigating, Thakkar2020UnderstandingUM, Thomas2020InvestigatingTI}, but instead of perturbing the training dataset, we search through the training datasets for the the gold solutions to the benchmarks.
Another line of work in studying memorization is to find documents related to the \textit{output} within a training dataset \cite{Lee2022DoLM, Peng2023NearDuplicateSS, Kandpal2022LargeLM, magar-schwartz-2022-data}. These works search the training dataset for documents relevant to an string and report the number of relevant documents. 
While we directly the effects of training data in model outputs, other approaches exist in using the model's weights to find the parts of the training dataset that influenced the model \cite{Han2022ORCAIP, Grosse2023StudyingLL}.
While this paper focuses on searching for contamination in open source models, many models are released without disclosing their training data. To search for contamination in these models, recent papers \cite{shi2023detecting, Oren2023proving, ranaldi2024investigating} use the probabilities of model outputs to observe contamination. This style of approach seems to work primarily when there are multiple copies within the training dataset, and is unreliable at detecting duplication rates of 1 \cite{Oren2023proving}. 
More recently, \citet{dong2024generalization} identifies contamination by measuring the peakness of model's output distribution via sampling, which works for black-box LLMs but provides less certainty compared with our method.

\noindent\textbf{Plagiarism detection.} Plagiarism detection is related to finding similar documents, and some work has already been done on evaluating the similarity of generated programs. \citet{pmlr-v202-yu23g} uses two methods, JPlag \cite{JPLAG} and Dolos \cite{maertens2022dolos} to calculate similarity scores between programs. Using the maximum score from the two methods, they determined any two programs with a similarity score greater than or equal to 0.5 to be potential plagiarism. Here, we only use Dolos, due to JPlag's restrictive license.

\section{Limitations}
\label{sec:limitations}

\noindent\textbf{Multiple correct solutions.} 
Due to the flexibility of programs, there can be multiple correct ways to solve a problem using Python. What we are searching for is only one possible solution presented as the gold solution, and as such we present our findings on the minimum number of questions to which models trained on the \pile and the \stack have been exposed to.

\noindent\textbf{Compute costs limiting search.}
Performing an $\mathcal{O}(n)$ search through the training corpus of different models is extremely costly. To reduce compute costs, we had to limit our search to relevant splits of the training corpus, which is the GitHub split for the \pile and the Python split for the \stack. It is possible for the models to see more of similar solutions in other parts of the pretraining corpus (\eg similar programs in Java for the \stack).

\noindent\textbf{False positives.}
While performing semantic-level comparison helps with the general recall of similar programs, 
it is possible for it to flag a program as being similar to the gold program despite being quite different. We present the results for all programs found to be perfect matches to the gold program in \autoref{sec:perfect_matches}, and some of examples show that false positives do exist. 

\noindent\textbf{Scarcity of open data.}
We only include two benchmarks and three models for this study. This is due to the scarcity of commonly used benchmarks that provide a gold program for every question, and models that have an open source training dataset. We hope more open models with open data will become available in the future to fuel further reasearch in this domain.

\section{Conclusion}
In this work, we attempt to quantify the data contamination issues for two popular code generation benchmarks, namely MBPP and HumanEval.
We use both surface-level and semantic similarity to exhaustively search among the pretraining data using gold solutions for these benchmarks.
By studying three series of models that are trained on two different corpus, the \pile and the \stack, we find that a significant portion of the benchmarks has solutions leaked into the pretraining data. Further analysis shows that models perform much better when similar solutions are seen during training, and such correlation is independent of the difficult and length of the problems.

\section*{Acknowledgements}

The authors would like to thank Rui Shen, Yilun Zhao and the late Dragomir Radev for their help and suggestions in early stages of the project. We would also like to thank Hailey Schoelkopf for helping us with the \pile dataset as well as helpful discussions and feedback.

\bibliography{anthology,custom}

\begin{thebibliography}{41}
\expandafter\ifx\csname natexlab\endcsname\relax\def\natexlab#1{#1}\fi

\bibitem[{Allamanis(2019)}]{10.1145/3359591.3359735}
Miltiadis Allamanis. 2019.
\newblock \href {https://doi.org/10.1145/3359591.3359735} {The adverse effects
  of code duplication in machine learning models of code}.
\newblock In \emph{Proceedings of the 2019 ACM SIGPLAN International Symposium
  on New Ideas, New Paradigms, and Reflections on Programming and Software},
  Onward! 2019, page 143–153, New York, NY, USA. Association for Computing
  Machinery.

\bibitem[{Austin et~al.(2021)Austin, Odena, Nye, Bosma, Michalewski, Dohan,
  Jiang, Cai, Terry, Le, and Sutton}]{austin2021program}
Jacob Austin, Augustus Odena, Maxwell Nye, Maarten Bosma, Henryk Michalewski,
  David Dohan, Ellen Jiang, Carrie Cai, Michael Terry, Quoc Le, and Charles
  Sutton. 2021.
\newblock \href {http://arxiv.org/abs/2108.07732} {Program synthesis with large
  language models}.

\bibitem[{Biderman et~al.(2023)Biderman, Schoelkopf, Anthony, Bradley, O'Brien,
  Hallahan, Khan, Purohit, Prashanth, Raff, Skowron, Sutawika, and Van
  Der~Wal}]{pmlr-v202-biderman23a}
Stella Biderman, Hailey Schoelkopf, Quentin~Gregory Anthony, Herbie Bradley,
  Kyle O'Brien, Eric Hallahan, Mohammad~Aflah Khan, Shivanshu Purohit,
  Usvsn~Sai Prashanth, Edward Raff, Aviya Skowron, Lintang Sutawika, and Oskar
  Van Der~Wal. 2023.
\newblock \href {https://proceedings.mlr.press/v202/biderman23a.html} {Pythia:
  A suite for analyzing large language models across training and scaling}.
\newblock In \emph{Proceedings of the 40th International Conference on Machine
  Learning}, volume 202 of \emph{Proceedings of Machine Learning Research},
  pages 2397--2430. PMLR.

\bibitem[{Blevins and Zettlemoyer(2022)}]{blevins-zettlemoyer-2022-language}
Terra Blevins and Luke Zettlemoyer. 2022.
\newblock \href {https://doi.org/10.18653/v1/2022.emnlp-main.233} {Language
  contamination helps explains the cross-lingual capabilities of {E}nglish
  pretrained models}.
\newblock In \emph{Proceedings of the 2022 Conference on Empirical Methods in
  Natural Language Processing}, pages 3563--3574, Abu Dhabi, United Arab
  Emirates. Association for Computational Linguistics.

\bibitem[{Carlini et~al.(2023)Carlini, Ippolito, Jagielski, Lee, Tramer, and
  Zhang}]{carlini2023quantifying}
Nicholas Carlini, Daphne Ippolito, Matthew Jagielski, Katherine Lee, Florian
  Tramer, and Chiyuan Zhang. 2023.
\newblock \href {http://arxiv.org/abs/2202.07646} {Quantifying memorization
  across neural language models}.

\bibitem[{Carlini et~al.(2020)Carlini, Tram{\`e}r, Wallace, Jagielski,
  Herbert-Voss, Lee, Roberts, Brown, Song, Erlingsson, Oprea, and
  Raffel}]{Carlini2020ExtractingTD}
Nicholas Carlini, Florian Tram{\`e}r, Eric Wallace, Matthew Jagielski, Ariel
  Herbert-Voss, Katherine Lee, Adam Roberts, Tom~B. Brown, Dawn~Xiaodong Song,
  {\'U}lfar Erlingsson, Alina Oprea, and Colin Raffel. 2020.
\newblock \href {https://api.semanticscholar.org/CorpusID:229156229}
  {Extracting training data from large language models}.
\newblock In \emph{USENIX Security Symposium}.

\bibitem[{Chang et~al.(2023)Chang, Cramer, Soni, and Bamman}]{chang2023speak}
Kent~K. Chang, Mackenzie Cramer, Sandeep Soni, and David Bamman. 2023.
\newblock \href {http://arxiv.org/abs/2305.00118} {Speak, memory: An
  archaeology of books known to chatgpt/gpt-4}.

\bibitem[{Chen et~al.(2021)Chen, Tworek, Jun, Yuan, de~Oliveira~Pinto, Kaplan,
  Edwards, Burda, Joseph, Brockman, Ray, Puri, Krueger, Petrov, Khlaaf, Sastry,
  Mishkin, Chan, Gray, Ryder, Pavlov, Power, Kaiser, Bavarian, Winter, Tillet,
  Such, Cummings, Plappert, Chantzis, Barnes, Herbert-Voss, Guss, Nichol,
  Paino, Tezak, Tang, Babuschkin, Balaji, Jain, Saunders, Hesse, Carr, Leike,
  Achiam, Misra, Morikawa, Radford, Knight, Brundage, Murati, Mayer, Welinder,
  McGrew, Amodei, McCandlish, Sutskever, and Zaremba}]{chen2021evaluating}
Mark Chen, Jerry Tworek, Heewoo Jun, Qiming Yuan, Henrique~Ponde
  de~Oliveira~Pinto, Jared Kaplan, Harri Edwards, Yuri Burda, Nicholas Joseph,
  Greg Brockman, Alex Ray, Raul Puri, Gretchen Krueger, Michael Petrov, Heidy
  Khlaaf, Girish Sastry, Pamela Mishkin, Brooke Chan, Scott Gray, Nick Ryder,
  Mikhail Pavlov, Alethea Power, Lukasz Kaiser, Mohammad Bavarian, Clemens
  Winter, Philippe Tillet, Felipe~Petroski Such, Dave Cummings, Matthias
  Plappert, Fotios Chantzis, Elizabeth Barnes, Ariel Herbert-Voss,
  William~Hebgen Guss, Alex Nichol, Alex Paino, Nikolas Tezak, Jie Tang, Igor
  Babuschkin, Suchir Balaji, Shantanu Jain, William Saunders, Christopher
  Hesse, Andrew~N. Carr, Jan Leike, Josh Achiam, Vedant Misra, Evan Morikawa,
  Alec Radford, Matthew Knight, Miles Brundage, Mira Murati, Katie Mayer, Peter
  Welinder, Bob McGrew, Dario Amodei, Sam McCandlish, Ilya Sutskever, and
  Wojciech Zaremba. 2021.
\newblock \href {http://arxiv.org/abs/2107.03374} {Evaluating large language
  models trained on code}.

\bibitem[{Chowdhery et~al.(2022)Chowdhery, Narang, Devlin, Bosma, Mishra,
  Roberts, Barham, Chung, Sutton, Gehrmann, Schuh, Shi, Tsvyashchenko, Maynez,
  Rao, Barnes, Tay, Shazeer, Prabhakaran, Reif, Du, Hutchinson, Pope, Bradbury,
  Austin, Isard, Gur-Ari, Yin, Duke, Levskaya, Ghemawat, Dev, Michalewski,
  Garcia, Misra, Robinson, Fedus, Zhou, Ippolito, Luan, Lim, Zoph, Spiridonov,
  Sepassi, Dohan, Agrawal, Omernick, Dai, Pillai, Pellat, Lewkowycz, Moreira,
  Child, Polozov, Lee, Zhou, Wang, Saeta, Diaz, Firat, Catasta, Wei,
  Meier-Hellstern, Eck, Dean, Petrov, and Fiedel}]{chowdhery2022palm}
Aakanksha Chowdhery, Sharan Narang, Jacob Devlin, Maarten Bosma, Gaurav Mishra,
  Adam Roberts, Paul Barham, Hyung~Won Chung, Charles Sutton, Sebastian
  Gehrmann, Parker Schuh, Kensen Shi, Sasha Tsvyashchenko, Joshua Maynez,
  Abhishek Rao, Parker Barnes, Yi~Tay, Noam Shazeer, Vinodkumar Prabhakaran,
  Emily Reif, Nan Du, Ben Hutchinson, Reiner Pope, James Bradbury, Jacob
  Austin, Michael Isard, Guy Gur-Ari, Pengcheng Yin, Toju Duke, Anselm
  Levskaya, Sanjay Ghemawat, Sunipa Dev, Henryk Michalewski, Xavier Garcia,
  Vedant Misra, Kevin Robinson, Liam Fedus, Denny Zhou, Daphne Ippolito, David
  Luan, Hyeontaek Lim, Barret Zoph, Alexander Spiridonov, Ryan Sepassi, David
  Dohan, Shivani Agrawal, Mark Omernick, Andrew~M. Dai,
  Thanumalayan~Sankaranarayana Pillai, Marie Pellat, Aitor Lewkowycz, Erica
  Moreira, Rewon Child, Oleksandr Polozov, Katherine Lee, Zongwei Zhou, Xuezhi
  Wang, Brennan Saeta, Mark Diaz, Orhan Firat, Michele Catasta, Jason Wei,
  Kathy Meier-Hellstern, Douglas Eck, Jeff Dean, Slav Petrov, and Noah Fiedel.
  2022.
\newblock \href {http://arxiv.org/abs/2204.02311} {Palm: Scaling language
  modeling with pathways}.

\bibitem[{Dodge et~al.(2021)Dodge, Sap, Marasovi{\'c}, Agnew, Ilharco,
  Groeneveld, Mitchell, and Gardner}]{dodge-etal-2021-documenting}
Jesse Dodge, Maarten Sap, Ana Marasovi{\'c}, William Agnew, Gabriel Ilharco,
  Dirk Groeneveld, Margaret Mitchell, and Matt Gardner. 2021.
\newblock \href {https://doi.org/10.18653/v1/2021.emnlp-main.98} {Documenting
  large webtext corpora: A case study on the colossal clean crawled corpus}.
\newblock In \emph{Proceedings of the 2021 Conference on Empirical Methods in
  Natural Language Processing}, pages 1286--1305, Online and Punta Cana,
  Dominican Republic. Association for Computational Linguistics.

\bibitem[{Dong et~al.(2024)Dong, Jiang, Liu, Jin, and
  Li}]{dong2024generalization}
Yihong Dong, Xue Jiang, Huanyu Liu, Zhi Jin, and Ge~Li. 2024.
\newblock Generalization or memorization: Data contamination and trustworthy
  evaluation for large language models.
\newblock \emph{arXiv preprint arXiv:2402.15938}.

\bibitem[{Gao et~al.(2020)Gao, Biderman, Black, Golding, Hoppe, Foster, Phang,
  He, Thite, Nabeshima, Presser, and Leahy}]{gao2020pile}
Leo Gao, Stella Biderman, Sid Black, Laurence Golding, Travis Hoppe, Charles
  Foster, Jason Phang, Horace He, Anish Thite, Noa Nabeshima, Shawn Presser,
  and Connor Leahy. 2020.
\newblock \href {http://arxiv.org/abs/2101.00027} {The pile: An 800gb dataset
  of diverse text for language modeling}.

\bibitem[{Golchin and Surdeanu(2023)}]{golchin2023time}
Shahriar Golchin and Mihai Surdeanu. 2023.
\newblock \href {http://arxiv.org/abs/2308.08493} {Time travel in llms: Tracing
  data contamination in large language models}.

\bibitem[{Grosse et~al.(2023)Grosse, Bae, Anil, Elhage, Tamkin, Tajdini,
  Steiner, Li, Durmus, Perez, Hubinger, Lukovsiut.e, Nguyen, Joseph,
  McCandlish, Kaplan, and Bowman}]{Grosse2023StudyingLL}
Roger~Baker Grosse, Juhan Bae, Cem Anil, Nelson Elhage, Alex Tamkin,
  Amirhossein Tajdini, Benoit Steiner, Dustin Li, Esin Durmus, Ethan Perez,
  Evan Hubinger, Kamil.e Lukovsiut.e, Karina Nguyen, Nicholas Joseph, Sam
  McCandlish, Jared Kaplan, and Sam Bowman. 2023.
\newblock \href {https://api.semanticscholar.org/CorpusID:260682872} {Studying
  large language model generalization with influence functions}.
\newblock \emph{ArXiv}, abs/2308.03296.

\bibitem[{Han and Tsvetkov(2022)}]{Han2022ORCAIP}
Xiaochuang Han and Yulia Tsvetkov. 2022.
\newblock \href {https://api.semanticscholar.org/CorpusID:249063011} {Orca:
  Interpreting prompted language models via locating supporting data evidence
  in the ocean of pretraining data}.
\newblock \emph{ArXiv}, abs/2205.12600.

\bibitem[{Henderson et~al.(2017)Henderson, Sinha, Angelard-Gontier, Ke, Fried,
  Lowe, and Pineau}]{Henderson2017EthicalCI}
Peter Henderson, Koustuv Sinha, Nicolas Angelard-Gontier, Nan~Rosemary Ke,
  Genevieve Fried, Ryan Lowe, and Joelle Pineau. 2017.
\newblock \href {https://api.semanticscholar.org/CorpusID:33499714} {Ethical
  challenges in data-driven dialogue systems}.
\newblock \emph{Proceedings of the 2018 AAAI/ACM Conference on AI, Ethics, and
  Society}.

\bibitem[{Hoffmann et~al.(2022)Hoffmann, Borgeaud, Mensch, Buchatskaya, Cai,
  Rutherford, de~Las~Casas, Hendricks, Welbl, Clark, Hennigan, Noland,
  Millican, van~den Driessche, Damoc, Guy, Osindero, Simonyan, Elsen, Rae,
  Vinyals, and Sifre}]{hoffmann2022training}
Jordan Hoffmann, Sebastian Borgeaud, Arthur Mensch, Elena Buchatskaya, Trevor
  Cai, Eliza Rutherford, Diego de~Las~Casas, Lisa~Anne Hendricks, Johannes
  Welbl, Aidan Clark, Tom Hennigan, Eric Noland, Katie Millican, George van~den
  Driessche, Bogdan Damoc, Aurelia Guy, Simon Osindero, Karen Simonyan, Erich
  Elsen, Jack~W. Rae, Oriol Vinyals, and Laurent Sifre. 2022.
\newblock \href {http://arxiv.org/abs/2203.15556} {Training compute-optimal
  large language models}.

\bibitem[{Jiang et~al.(2024)Jiang, Liu, Zhong, Schaeffer, Ouyang, Han, and
  Koyejo}]{jiang2024investigating}
Minhao Jiang, Ken~Ziyu Liu, Ming Zhong, Rylan Schaeffer, Siru Ouyang, Jiawei
  Han, and Sanmi Koyejo. 2024.
\newblock \href {http://arxiv.org/abs/2401.06059} {Investigating data
  contamination for pre-training language models}.

\bibitem[{Kandpal et~al.(2022{\natexlab{a}})Kandpal, Deng, Roberts, Wallace,
  and Raffel}]{Kandpal2022LargeLM}
Nikhil Kandpal, H.~Deng, Adam Roberts, Eric Wallace, and Colin Raffel.
  2022{\natexlab{a}}.
\newblock \href {https://api.semanticscholar.org/CorpusID:253522998} {Large
  language models struggle to learn long-tail knowledge}.
\newblock In \emph{International Conference on Machine Learning}.

\bibitem[{Kandpal et~al.(2022{\natexlab{b}})Kandpal, Wallace, and
  Raffel}]{kandpal2022deduplicating}
Nikhil Kandpal, Eric Wallace, and Colin Raffel. 2022{\natexlab{b}}.
\newblock \href {http://arxiv.org/abs/2202.06539} {Deduplicating training data
  mitigates privacy risks in language models}.

\bibitem[{Kaplan et~al.(2020)Kaplan, McCandlish, Henighan, Brown, Chess, Child,
  Gray, Radford, Wu, and Amodei}]{kaplan2020scaling}
Jared Kaplan, Sam McCandlish, Tom Henighan, Tom~B. Brown, Benjamin Chess, Rewon
  Child, Scott Gray, Alec Radford, Jeffrey Wu, and Dario Amodei. 2020.
\newblock \href {http://arxiv.org/abs/2001.08361} {Scaling laws for neural
  language models}.

\bibitem[{Karmakar et~al.(2022)Karmakar, Prenner, D'Ambros, and
  Robbes}]{karmakar2022codex}
Anjan Karmakar, Julian~Aron Prenner, Marco D'Ambros, and Romain Robbes. 2022.
\newblock \href {http://arxiv.org/abs/2212.02684} {Codex hacks hackerrank:
  Memorization issues and a framework for code synthesis evaluation}.

\bibitem[{Kocetkov et~al.(2022)Kocetkov, Li, Allal, Li, Mou, Ferrandis,
  Jernite, Mitchell, Hughes, Wolf, Bahdanau, von Werra, and
  de~Vries}]{kocetkov2022stack}
Denis Kocetkov, Raymond Li, Loubna~Ben Allal, Jia Li, Chenghao Mou,
  Carlos~Muñoz Ferrandis, Yacine Jernite, Margaret Mitchell, Sean Hughes,
  Thomas Wolf, Dzmitry Bahdanau, Leandro von Werra, and Harm de~Vries. 2022.
\newblock \href {http://arxiv.org/abs/2211.15533} {The stack: 3 tb of
  permissively licensed source code}.

\bibitem[{Lee et~al.(2022)Lee, Le, Chen, and Lee}]{Lee2022DoLM}
Jooyoung Lee, Thai Le, Jinghui Chen, and Dongwon Lee. 2022.
\newblock \href {https://api.semanticscholar.org/CorpusID:247450984} {Do
  language models plagiarize?}
\newblock \emph{Proceedings of the ACM Web Conference 2023}.

\bibitem[{Levenshtein(1965)}]{Levenshtein1965BinaryCC}
Vladimir~I. Levenshtein. 1965.
\newblock \href {https://api.semanticscholar.org/CorpusID:60827152} {Binary
  codes capable of correcting deletions, insertions, and reversals}.
\newblock \emph{Soviet physics. Doklady}, 10:707--710.

\bibitem[{Li et~al.(2023)Li, Allal, Zi, Muennighoff, Kocetkov, Mou, Marone,
  Akiki, Li, Chim, Liu, Zheltonozhskii, Zhuo, Wang, Dehaene, Davaadorj,
  Lamy-Poirier, Monteiro, Shliazhko, Gontier, Meade, Zebaze, Yee, Umapathi,
  Zhu, Lipkin, Oblokulov, Wang, Murthy, Stillerman, Patel, Abulkhanov, Zocca,
  Dey, Zhang, Fahmy, Bhattacharyya, Yu, Singh, Luccioni, Villegas, Kunakov,
  Zhdanov, Romero, Lee, Timor, Ding, Schlesinger, Schoelkopf, Ebert, Dao,
  Mishra, Gu, Robinson, Anderson, Dolan-Gavitt, Contractor, Reddy, Fried,
  Bahdanau, Jernite, Ferrandis, Hughes, Wolf, Guha, von Werra, and
  de~Vries}]{li2023starcoder}
Raymond Li, Loubna~Ben Allal, Yangtian Zi, Niklas Muennighoff, Denis Kocetkov,
  Chenghao Mou, Marc Marone, Christopher Akiki, Jia Li, Jenny Chim, Qian Liu,
  Evgenii Zheltonozhskii, Terry~Yue Zhuo, Thomas Wang, Olivier Dehaene, Mishig
  Davaadorj, Joel Lamy-Poirier, João Monteiro, Oleh Shliazhko, Nicolas
  Gontier, Nicholas Meade, Armel Zebaze, Ming-Ho Yee, Logesh~Kumar Umapathi,
  Jian Zhu, Benjamin Lipkin, Muhtasham Oblokulov, Zhiruo Wang, Rudra Murthy,
  Jason Stillerman, Siva~Sankalp Patel, Dmitry Abulkhanov, Marco Zocca, Manan
  Dey, Zhihan Zhang, Nour Fahmy, Urvashi Bhattacharyya, Wenhao Yu, Swayam
  Singh, Sasha Luccioni, Paulo Villegas, Maxim Kunakov, Fedor Zhdanov, Manuel
  Romero, Tony Lee, Nadav Timor, Jennifer Ding, Claire Schlesinger, Hailey
  Schoelkopf, Jan Ebert, Tri Dao, Mayank Mishra, Alex Gu, Jennifer Robinson,
  Carolyn~Jane Anderson, Brendan Dolan-Gavitt, Danish Contractor, Siva Reddy,
  Daniel Fried, Dzmitry Bahdanau, Yacine Jernite, Carlos~Muñoz Ferrandis, Sean
  Hughes, Thomas Wolf, Arjun Guha, Leandro von Werra, and Harm de~Vries. 2023.
\newblock \href {http://arxiv.org/abs/2305.06161} {Starcoder: may the source be
  with you!}

\bibitem[{Maertens et~al.(2022)Maertens, Van~Petegem, Strijbol, Baeyens,
  Jacobs, Dawyndt, and Mesuere}]{maertens2022dolos}
Rien Maertens, Charlotte Van~Petegem, Niko Strijbol, Toon Baeyens, Arne~Carla
  Jacobs, Peter Dawyndt, and Bart Mesuere. 2022.
\newblock Dolos: Language-agnostic plagiarism detection in source code.
\newblock \emph{Journal of Computer Assisted Learning}, 38(4):1046--1061.

\bibitem[{Magar and Schwartz(2022)}]{magar-schwartz-2022-data}
Inbal Magar and Roy Schwartz. 2022.
\newblock \href {https://doi.org/10.18653/v1/2022.acl-short.18} {Data
  contamination: From memorization to exploitation}.
\newblock In \emph{Proceedings of the 60th Annual Meeting of the Association
  for Computational Linguistics (Volume 2: Short Papers)}, pages 157--165,
  Dublin, Ireland. Association for Computational Linguistics.

\bibitem[{Ni et~al.(2023)Ni, Yin, Zhao, Riddell, Feng, Shen, Yin, Liu, Yavuz,
  Xiong et~al.}]{ni2023l2ceval}
Ansong Ni, Pengcheng Yin, Yilun Zhao, Martin Riddell, Troy Feng, Rui Shen,
  Stephen Yin, Ye~Liu, Semih Yavuz, Caiming Xiong, et~al. 2023.
\newblock L2ceval: Evaluating language-to-code generation capabilities of large
  language models.
\newblock \emph{arXiv preprint arXiv:2309.17446}.

\bibitem[{Nijkamp et~al.(2023)Nijkamp, Pang, Hayashi, Tu, Wang, Zhou, Savarese,
  and Xiong}]{nijkamp2023CodeGen}
Erik Nijkamp, Bo~Pang, Hiroaki Hayashi, Lifu Tu, Huan Wang, Yingbo Zhou, Silvio
  Savarese, and Caiming Xiong. 2023.
\newblock \href {http://arxiv.org/abs/2203.13474} {Codegen: An open large
  language model for code with multi-turn program synthesis}.

\bibitem[{Oren and Meister(2023)}]{Oren2023proving}
Yonatan Oren and Nicole Meister. 2023.
\newblock \href {http://arxiv.org/abs/2310.17623} {Proving test set
  contamination in black box language models}.

\bibitem[{Peng et~al.(2023)Peng, Wang, and Deng}]{Peng2023NearDuplicateSS}
Zhen Peng, Zhizhi Wang, and Dong Deng. 2023.
\newblock \href {https://api.semanticscholar.org/CorpusID:259213212}
  {Near-duplicate sequence search at scale for large language model
  memorization evaluation}.
\newblock \emph{Proceedings of the ACM on Management of Data}, 1:1 -- 18.

\bibitem[{Prechelt and Malpohl(2003)}]{JPLAG}
Lutz Prechelt and Guido Malpohl. 2003.
\newblock Finding plagiarisms among a set of programs with jplag.
\newblock \emph{Journal of Universal Computer Science}, 8.

\bibitem[{Ranaldi et~al.(2024)Ranaldi, Ruzzetti, Onorati, Ranaldi, Giannone,
  Favalli, Romagnoli, and Zanzotto}]{ranaldi2024investigating}
Federico Ranaldi, Elena~Sofia Ruzzetti, Dario Onorati, Leonardo Ranaldi,
  Cristina Giannone, Andrea Favalli, Raniero Romagnoli, and Fabio~Massimo
  Zanzotto. 2024.
\newblock \href {http://arxiv.org/abs/2402.08100} {Investigating the impact of
  data contamination of large language models in text-to-sql translation}.

\bibitem[{Razeghi et~al.(2022)Razeghi, Logan~IV, Gardner, and
  Singh}]{razeghi-etal-2022-impact}
Yasaman Razeghi, Robert~L Logan~IV, Matt Gardner, and Sameer Singh. 2022.
\newblock \href {https://doi.org/10.18653/v1/2022.findings-emnlp.59} {Impact of
  pretraining term frequencies on few-shot numerical reasoning}.
\newblock In \emph{Findings of the Association for Computational Linguistics:
  EMNLP 2022}, pages 840--854, Abu Dhabi, United Arab Emirates. Association for
  Computational Linguistics.

\bibitem[{Sarkar et~al.(2016)Sarkar, Das, Pakray, and
  Gelbukh}]{sarkar-etal-2016-junitmz}
Sandip Sarkar, Dipankar Das, Partha Pakray, and Alexander Gelbukh. 2016.
\newblock \href {https://doi.org/10.18653/v1/S16-1108} {{JUNITMZ} at
  {S}em{E}val-2016 task 1: Identifying semantic similarity using {L}evenshtein
  ratio}.
\newblock In \emph{Proceedings of the 10th International Workshop on Semantic
  Evaluation ({S}em{E}val-2016)}, pages 702--705, San Diego, California.
  Association for Computational Linguistics.

\bibitem[{Shi et~al.(2023)Shi, Ajith, Xia, Huang, Liu, Blevins, Chen, and
  Zettlemoyer}]{shi2023detecting}
Weijia Shi, Anirudh Ajith, Mengzhou Xia, Yangsibo Huang, Daogao Liu, Terra
  Blevins, Danqi Chen, and Luke Zettlemoyer. 2023.
\newblock \href {http://arxiv.org/abs/2310.16789} {Detecting pretraining data
  from large language models}.

\bibitem[{Thakkar et~al.(2020)Thakkar, Ramaswamy, Mathews, and
  Beaufays}]{Thakkar2020UnderstandingUM}
Om~Thakkar, Swaroop~Indra Ramaswamy, Rajiv Mathews, and Franccoise Beaufays.
  2020.
\newblock \href {https://api.semanticscholar.org/CorpusID:219686873}
  {Understanding unintended memorization in federated learning}.
\newblock \emph{ArXiv}, abs/2006.07490.

\bibitem[{Thomas et~al.(2020)Thomas, Adelani, Davody, Mogadala, and
  Klakow}]{Thomas2020InvestigatingTI}
Aleena~Anna Thomas, David~Ifeoluwa Adelani, Ali Davody, Aditya Mogadala, and
  Dietrich Klakow. 2020.
\newblock \href {https://api.semanticscholar.org/CorpusID:220658693}
  {Investigating the impact of pre-trained word embeddings on memorization in
  neural networks}.
\newblock In \emph{Workshop on Time-Delay Systems}.

\bibitem[{Yang et~al.(2023)Yang, Chiang, Zheng, Gonzalez, and
  Stoica}]{yang2023rethinking}
Shuo Yang, Wei-Lin Chiang, Lianmin Zheng, Joseph~E. Gonzalez, and Ion Stoica.
  2023.
\newblock \href {http://arxiv.org/abs/2311.04850} {Rethinking benchmark and
  contamination for language models with rephrased samples}.

\bibitem[{Yu et~al.(2023)Yu, Wu, Zhang, Wang, Vorobeychik, and
  Xiao}]{pmlr-v202-yu23g}
Zhiyuan Yu, Yuhao Wu, Ning Zhang, Chenguang Wang, Yevgeniy Vorobeychik, and
  Chaowei Xiao. 2023.
\newblock \href {https://proceedings.mlr.press/v202/yu23g.html}
  {{C}ode{IPP}rompt: Intellectual property infringement assessment of code
  language models}.
\newblock In \emph{Proceedings of the 40th International Conference on Machine
  Learning}, volume 202 of \emph{Proceedings of Machine Learning Research},
  pages 40373--40389. PMLR.

\end{thebibliography}
\bibliographystyle{acl_natbib}

\newpage

\appendix

\newpage
\section{Additional Examples and Results}
\label{sec:additional-examples}

\subsection{All Model Series}
In \autoref{tab:performance_gap_both} we present the results on the largest versions of the StarCoder-Base, Pythia and CodeGen-nl model series. In \autoref{tab:performance_gap_full} we show results on four model versions from each model series.

\subsection{Relevant Info for Models on the HumanEval Benchmark}
\begin{figure}[H]
  \centering
  \begin{subfigure}[b]{.22\textwidth}
    \includegraphics[scale=.45]{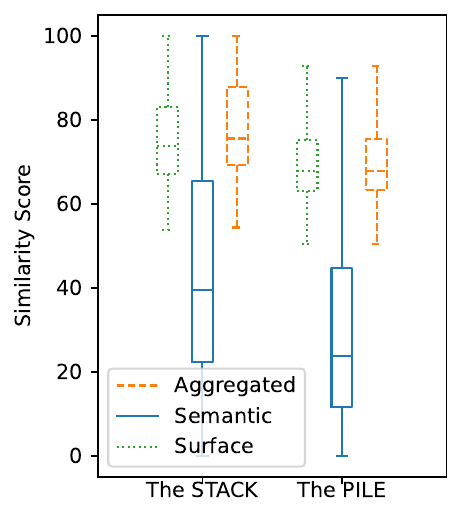}
    \caption{Top-10 Scores.}
    \label{fig:sub1}
  \end{subfigure}
  \begin{subfigure}[b]{.22\textwidth}
    \includegraphics[scale=.45]{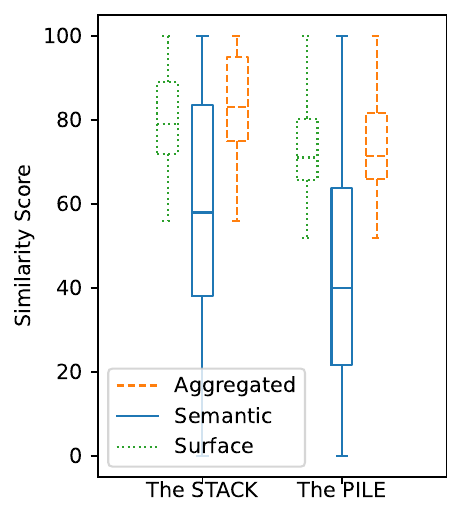}
    \caption{Top-1 Score.}
    \label{fig:sub2}
  \end{subfigure}
  \caption{We show the similarity scores for both The \pile and The \textsc{Stack} found by searching for answers to the gold programs in the HumanEval benchmark. We compare the similarity scores from different techniques, as well as the difference between using the top-1 score and the top-10 scores.
  }
  \label{fig:HE_relevant_score_info}
\end{figure}

\begin{figure}[H]
    \centering
    \includegraphics[scale=.5]{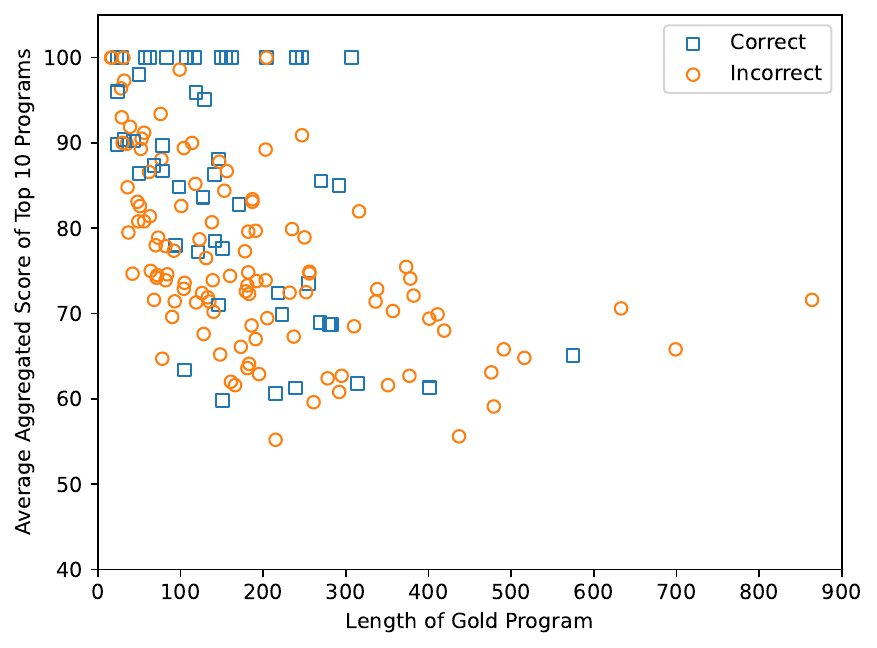}
    \caption{Length of gold programs in the HumanEval benchmark plotted against the average aggregated similarity score of the top-10 scores within the training dataset used for the StarCoderBase model family.}
    \label{fig:he_size_against_similarity}
\end{figure}

We provide versions of \autoref{fig:relevant_score_info} and \autoref{fig:size_against_similarity} for the HumanEval benchmark in  \autoref{fig:HE_relevant_score_info} and \autoref{fig:he_size_against_similarity} respectively.

\subsection{Examples of Similarity Scores}
\label{sec:examples-of-similarity-scores}
In \autoref{fig:dolos_score_example} and \autoref{fig:levenshtein_score_example} we provide examples of programs found within the training data and the relevant similarity score returned for them. A similarity score of 70 typically represents a program that is no longer similar to the gold program.

\begin{figure}[H]
    \centering
    \include{tabs/dolos_examples}
    \vspace*{-5mm}\caption{Examples of different programs and their corresponding Dolos scores when compared to a gold program from the MBPP benchmark.}
    \label{fig:dolos_score_example}
\end{figure}

\begin{figure}[H]
    \centering
    \include{tabs/levensthein_examples}
    \vspace*{-5mm}\caption{Examples of different programs and their corresponding Levenshtein scores when compared to a gold program from the MBPP benchmark.}
    \label{fig:levenshtein_score_example}
\end{figure}

\begin{table*}[t]
\small
\centering
\begin{tabular}{lrrrrrrr}
\toprule
\multicolumn{1}{c}{\multirow{2}{*}{\textbf{Models}}} &  \multicolumn{3}{c}{\textbf{MBPP}}                                                      & \multicolumn{1}{c}{} & \multicolumn{3}{c}{\textbf{HumanEval}}                                                 \\ \cline{2-4} \cline{6-8} 
 & \multicolumn{1}{c}{$Acc_o$} & \multicolumn{1}{c}{$\Uparrow^{10\%}$} & \multicolumn{1}{c}{$\Downarrow_{10\%}$ ($\Delta_\Updownarrow$)} & \multicolumn{1}{c}{} & \multicolumn{1}{c}{$Acc_o$} & \multicolumn{1}{c}{$\Uparrow^{10\%}$} & \multicolumn{1}{c}{$\Downarrow_{10\%}$ ($\Delta_\Updownarrow$)} \\ \midrule
StarCoderBase-1B    & 23.4                    & 54.0                    & 4.0 (-50.0)                      &                      & 16.5                    & 56.6                    & 18.8 (-37.5)                     \\
StarCoderBase-3B    & 29.8                    & 64.0                    & 8.0 (-56.0)                      &                      & 22.0                    & 75.0                    & 25.0 (-50.0)                     \\
StarCoderBase-7B    & 37.2                    & 70.0                    & 18.0 (-52.0)                     &                      & 31.3                    & 68.8                    & 31.3 (-37.5)                     \\
StarCoderBase-15.5B & 41.6                    & 72.0                    & 22.0 (-50.0)                     &                      & 30.5                    & 75.0                    & 31.3 (-43.7)                     \\ \midrule
Pythia-1.4B         & 4.4                     & 18.0                    & 0.0 (-18.0)                      &                      & 4.9                     & 25.0                    & 0.0 (-25.0)                      \\
Pythia-2.8B         & 12.0                    & 30.0                    & 2.0 (-28.0)                      &                      & 7.3                     & 31.3                    & 0.0 (-31.3)                      \\
Pythia-6.9B         & 12.6                    & 34.0                    & 2.0 (-32.0)                      &                      & 6.7                     & 43.8                    & 0.0 (-43.8)                      \\
Pythia-12B          & 17.8                    & 40.0                    & 8.0 (-32.0)                      &                      & 9.8                     & 56.3                    & 0.0 (-56.3)                      \\ \midrule
CodeGen-NL-350M     & 2.2                     & 8.0                     & 0.0 (-8.0)                       &                      & 3.0                     & 12.5                    & 0.0 (-12.5)                      \\
CodeGen-NL-2B       & 12.7                    & 36.0                    & 4.0 (-32.0)                      &                      & 7.9                     & 37.5                    & 0.0 (-37.5)                      \\
CodeGen-NL-6B       & 15.8                    & 42.0                    & 6.0 (-36.0)                      &                      & 8.5                     & 37.5                    & 0.0 (-37.5)                      \\
CodeGen-NL-16B      & 19.6                    & 48.0                    & 6.0 (-42.0)                      &                      & 14.6                    & 62.5                    & 0.0 (-62.5)                      \\ \bottomrule
\end{tabular}
\caption{We show the performance gap ($\Delta_\Updownarrow$) between the top 10\% ($\Uparrow^{10\%}$) and bottom 10\% ($\Downarrow_{10\%}$) of questions for the MBPP and HumanEval benchmarks compared against "$Acc_o$" the original model accuracy. This is the full version of \autoref{tab:performance_gap_both}, showing results for four models in each model series.
}
\label{tab:performance_gap_full}
\end{table*}

\subsection{Perfect Matches}
\label{sec:perfect_matches}

We provide lists of every question that was seen 10 or more times during training for models trained on a specific dataset. These can be found in \autoref{tab:seen-start} through \autoref{tab:seen-end}.

\begingroup

\setlength{\tabcolsep}{6pt} % Default value: 6pt
\renewcommand{\arraystretch}{0} % Default value: 1

\begin{table*}[]
\small
\begin{tabular}{L{4cm}|L{4cm}|L{3cm}}
\toprule
\textbf{Natural Language Question} & \textbf{Gold Program}                                                              & \textbf{Found 100\% Matches}                                                       \\ \midrule 
\lstset{linewidth=5.5cm, frame=none, numbers=none, backgroundcolor=\color{white}, language=Python}
\begin{lstlisting}
Write a function to find the closest smaller number than n.
\end{lstlisting}
&
\lstset{linewidth=5.5cm, frame=none, numbers=none, backgroundcolor=\color{white}, language=Python}
\begin{lstlisting}
def closest_num(N):
  return (N - 1)
\end{lstlisting}
&
{\lstset{linewidth=3.5cm}
\begin{lstlisting}[frame=none, numbers=none, backgroundcolor=\color{white}][language=Python]
def parent(i):
    return (i - 1)\end{lstlisting}}   \\ \midrule 
{\lstset{linewidth=5.5cm}
\begin{lstlisting}[frame=none, numbers=none, backgroundcolor=\color{white}][language=Python]
Write a python function to count true booleans in the given list.\end{lstlisting}}   &
{\lstset{linewidth=5.5cm}
\begin{lstlisting}[frame=none, numbers=none, backgroundcolor=\color{white}][language=Python]
def count(lst):   
    return sum(lst)\end{lstlisting}}   &
{\lstset{linewidth=3.5cm}
\begin{lstlisting}[frame=none, numbers=none, backgroundcolor=\color{white}][language=Python]

def average(lst):
    return sum(lst) \end{lstlisting}}   \\ \midrule 
{\lstset{linewidth=5.5cm}
\begin{lstlisting}[frame=none, numbers=none, backgroundcolor=\color{white}][language=Python]
Write a python function to find smallest number in a list.\end{lstlisting}}   &
{\lstset{linewidth=5.5cm}
\begin{lstlisting}[frame=none, numbers=none, backgroundcolor=\color{white}][language=Python]
def smallest_num(xs):
  return min(xs)\end{lstlisting}}   &
{\lstset{linewidth=3.5cm}
\begin{lstlisting}[frame=none, numbers=none, backgroundcolor=\color{white}][language=Python]
def min_usecase3(x):
    return min(x)\end{lstlisting}}   \\ \bottomrule
\end{tabular}
\caption{All questions flagged as being seen by models trained on the \pile 10 or more times within the MBPP benchmark}
\label{tab:seen-start}
\end{table*}

\endgroup
\begingroup

\setlength{\tabcolsep}{6pt} % Default value: 6pt
\renewcommand{\arraystretch}{0} % Default value: 1

\begin{table*}[]
\small
\begin{tabular}{L{4cm}|L{4cm}|L{3cm}}
\toprule
\textbf{Natural Language Question} & \textbf{Gold Program}                                                              & \textbf{Found 100\% Matches}                                                       \\ \midrule 
{\lstset{linewidth=5.5cm}
\begin{lstlisting}[frame=none, numbers=none, backgroundcolor=\color{white}][language=Python]
Write a function to find the n-th rectangular number.\end{lstlisting}}   &
{\lstset{linewidth=5.5cm}
\begin{lstlisting}[frame=none, numbers=none, backgroundcolor=\color{white}][language=Python]
def find_rect_num(n):
  return n*(n + 1) \end{lstlisting}}   &
{\lstset{linewidth=3.5cm}
\begin{lstlisting}[frame=none, numbers=none, backgroundcolor=\color{white}][language=Python]


def get_sum(n):
    return n * (n + 1) \end{lstlisting}}   \\ \midrule 
{\lstset{linewidth=5.5cm}
\begin{lstlisting}[frame=none, numbers=none, backgroundcolor=\color{white}][language=Python]
Write a python function to find the last digit of a given number.\end{lstlisting}}   &
{\lstset{linewidth=5.5cm}
\begin{lstlisting}[frame=none, numbers=none, backgroundcolor=\color{white}][language=Python]
def last_Digit(n) :
    return (n % 10) \end{lstlisting}}   &
{\lstset{linewidth=3.5cm}
\begin{lstlisting}[frame=none, numbers=none, backgroundcolor=\color{white}][language=Python]
def shift_right(b):
    return (b << 1) \end{lstlisting}}   \\ \midrule 
{\lstset{linewidth=5.5cm}
\begin{lstlisting}[frame=none, numbers=none, backgroundcolor=\color{white}][language=Python]
Write a function to find the closest smaller number than n.\end{lstlisting}}   &
{\lstset{linewidth=5.5cm}
\begin{lstlisting}[frame=none, numbers=none, backgroundcolor=\color{white}][language=Python]
def closest_num(N):
  return (N - 1)\end{lstlisting}}   &
{\lstset{linewidth=3.5cm}
\begin{lstlisting}[frame=none, numbers=none, backgroundcolor=\color{white}][language=Python]

def percentage(x):
  return (x - 1)\end{lstlisting}}   \\ \midrule 
{\lstset{linewidth=5.5cm}
\begin{lstlisting}[frame=none, numbers=none, backgroundcolor=\color{white}][language=Python]
Write a python function to find smallest number in a list.\end{lstlisting}}   &
{\lstset{linewidth=5.5cm}
\begin{lstlisting}[frame=none, numbers=none, backgroundcolor=\color{white}][language=Python]
def smallest_num(xs):
  return min(xs)\end{lstlisting}}   &
{\lstset{linewidth=3.5cm}
\begin{lstlisting}[frame=none, numbers=none, backgroundcolor=\color{white}][language=Python]

def smallest(l):
    return min(l)
\end{lstlisting}}   \\ \midrule 
{\lstset{linewidth=5.5cm}
\begin{lstlisting}[frame=none, numbers=none, backgroundcolor=\color{white}][language=Python]
Write a python function to count positive numbers in a list.\end{lstlisting}}   &
{\lstset{linewidth=5.5cm}
\begin{lstlisting}[frame=none, numbers=none, backgroundcolor=\color{white}][language=Python]
def pos_count(list):
  pos_count= 0
  for num in list: 
    if num >= 0: 
      pos_count += 1
  return pos_count \end{lstlisting}}   &
{\lstset{linewidth=3.5cm}
\begin{lstlisting}[frame=none, numbers=none, backgroundcolor=\color{white}][language=Python]
def array_count9(nums):
    count = 0
    for num in nums:
      if num == 9:
        count += 1
    return count \end{lstlisting}}   \\ \midrule 
{\lstset{linewidth=5.5cm}
\begin{lstlisting}[frame=none, numbers=none, backgroundcolor=\color{white}][language=Python]
Write a function to swap two numbers.\end{lstlisting}}   &
{\lstset{linewidth=5.5cm}
\begin{lstlisting}[frame=none, numbers=none, backgroundcolor=\color{white}][language=Python]
def swap_numbers(a,b):
 temp = a
 a = b
 b = temp
 return (a,b)\end{lstlisting}}   &
{\lstset{linewidth=3.5cm}
\begin{lstlisting}[frame=none, numbers=none, backgroundcolor=\color{white}][language=Python]
def swap(a,b):
    tmp = a
    a = b
    b = tmp
    return a,b\end{lstlisting}}   \\ \midrule 
{\lstset{linewidth=5.5cm}
\begin{lstlisting}[frame=none, numbers=none, backgroundcolor=\color{white}][language=Python]
[link text](https:// [link text](https:// [link text](https://)))write a function to convert a string to a list.
\end{lstlisting}}   &
{\lstset{linewidth=5.5cm}
\begin{lstlisting}[frame=none, numbers=none, backgroundcolor=\color{white}][language=Python]
def string_to_list(string): 
    lst = list(string.split(" ")) 
    return lst\end{lstlisting}}   &
{\lstset{linewidth=3.5cm}
\begin{lstlisting}[frame=none, numbers=none, backgroundcolor=\color{white}][language=Python]

def String_to_list (Strings):
    list1=list(Strings.split(" "))
    return l\end{lstlisting}}   \\ \midrule 
{\lstset{linewidth=5.5cm}
\begin{lstlisting}[frame=none, numbers=none, backgroundcolor=\color{white}][language=Python]
Write a python function to find the minimum of two numbers.\end{lstlisting}}   &
{\lstset{linewidth=5.5cm}
\begin{lstlisting}[frame=none, numbers=none, backgroundcolor=\color{white}][language=Python]
def minimum(a,b):   
    if a <= b: 
        return a 
    else: 
        return b \end{lstlisting}}   &
{\lstset{linewidth=3.5cm}
\begin{lstlisting}[frame=none, numbers=none, backgroundcolor=\color{white}][language=Python]
def minimum(a, b):
    if a <= b:
        return a
    else:
        return b

def \end{lstlisting}}   \\ \midrule 
{\lstset{linewidth=5.5cm}
\begin{lstlisting}[frame=none, numbers=none, backgroundcolor=\color{white}][language=Python]
Write a function to check whether an element exists within a tuple.\end{lstlisting}}   &
{\lstset{linewidth=5.5cm}
\begin{lstlisting}[frame=none, numbers=none, backgroundcolor=\color{white}][language=Python]
def check_tuplex(tuplex,tuple1): 
  if tuple1 in tuplex:
    return True
  else:
     return False\end{lstlisting}}   &
{\lstset{linewidth=3.5cm}
\begin{lstlisting}[frame=none, numbers=none, backgroundcolor=\color{white}][language=Python]
def check_for_tag(ele, tag):
    if tag in ele:
        return True
    else:
        return False\end{lstlisting}}   \\ \midrule 
{\lstset{linewidth=5.5cm}
\begin{lstlisting}[frame=none, numbers=none, backgroundcolor=\color{white}][language=Python]
Write a python function to count true booleans in the given list.\end{lstlisting}}   &
{\lstset{linewidth=5.5cm}
\begin{lstlisting}[frame=none, numbers=none, backgroundcolor=\color{white}][language=Python]
def count(lst):   
    return sum(lst) \end{lstlisting}}   &
{\lstset{linewidth=3.5cm}
\begin{lstlisting}[frame=none, numbers=none, backgroundcolor=\color{white}][language=Python]
def mean(lst):
        return sum(lst) \end{lstlisting}}   \\ \bottomrule
\end{tabular}
\caption{All questions flagged as being seen by models trained on the \stack 10 or more times within the MBPP benchmark (Part 1)}
\end{table*}

\endgroup
\begingroup

\setlength{\tabcolsep}{6pt} % Default value: 6pt
\renewcommand{\arraystretch}{0} % Default value: 1

\begin{table*}[]
\small
\begin{tabular}{L{4cm}|L{4cm}|L{3cm}}
\toprule
\textbf{Natural Language Question} & \textbf{Gold Program}                                                              & \textbf{Found 100\% Matches}                                                       \\ \midrule 
{\lstset{linewidth=5.5cm}
\begin{lstlisting}[frame=none, numbers=none, backgroundcolor=\color{white}][language=Python]
Write a python function to find the largest negative number from the given list.\end{lstlisting}}   &
{\lstset{linewidth=5.5cm}
\begin{lstlisting}[frame=none, numbers=none, backgroundcolor=\color{white}][language=Python]
def largest_neg(list1): 
    max = list1[0] 
    for x in list1: 
        if x < max : 
             max = x  
    return max\end{lstlisting}}   &
{\lstset{linewidth=3.5cm}
\begin{lstlisting}[frame=none, numbers=none, backgroundcolor=\color{white}][language=Python]
def minimum( list ):
    min = list[ 0 ]
    for i in list:
        if i < min:
            min = i
    return min\end{lstlisting}}   \\ \midrule 
{\lstset{linewidth=5.5cm}
\begin{lstlisting}[frame=none, numbers=none, backgroundcolor=\color{white}][language=Python]
Write a function to convert the given decimal number to its binary equivalent.\end{lstlisting}}   &
{\lstset{linewidth=5.5cm}
\begin{lstlisting}[frame=none, numbers=none, backgroundcolor=\color{white}][language=Python]
def decimal_to_binary(n): 
    return bin(n).replace("0b","") \end{lstlisting}}   &
{\lstset{linewidth=3.5cm}
\begin{lstlisting}[frame=none, numbers=none, backgroundcolor=\color{white}][language=Python]

def decimal_to_binary(n):
    return bin(n).replace("0b", "")\end{lstlisting}}   \\ \midrule 
{\lstset{linewidth=5.5cm}
\begin{lstlisting}[frame=none, numbers=none, backgroundcolor=\color{white}][language=Python]
Write a python function to find the maximum of two numbers.\end{lstlisting}}   &
{\lstset{linewidth=5.5cm}
\begin{lstlisting}[frame=none, numbers=none, backgroundcolor=\color{white}][language=Python]
def maximum(a,b):   
    if a >= b: 
        return a 
    else: 
        return b \end{lstlisting}}   &
{\lstset{linewidth=3.5cm}
\begin{lstlisting}[frame=none, numbers=none, backgroundcolor=\color{white}][language=Python]
def maximum(a, b):
     
    if a >= b:
        return a
    else:
        return b\end{lstlisting}}   \\ \midrule 
{\lstset{linewidth=5.5cm}
\begin{lstlisting}[frame=none, numbers=none, backgroundcolor=\color{white}][language=Python]
Write a function to extract every specified element from a given two dimensional list.\end{lstlisting}}   &
{\lstset{linewidth=5.5cm}
\begin{lstlisting}[frame=none, numbers=none, backgroundcolor=\color{white}][language=Python]
def specified_element(nums, N):
    result = [i[N] for i in nums]
    return result\end{lstlisting}}   &
{\lstset{linewidth=3.5cm}
\begin{lstlisting}[frame=none, numbers=none, backgroundcolor=\color{white}][language=Python]

def _get_mean(names, table):
    x = [table[name] for name in names]
    return su\end{lstlisting}}   \\ \midrule 
{\lstset{linewidth=5.5cm}
\begin{lstlisting}[frame=none, numbers=none, backgroundcolor=\color{white}][language=Python]
Write a function to find the nth hexagonal number.\end{lstlisting}}   &
{\lstset{linewidth=5.5cm}
\begin{lstlisting}[frame=none, numbers=none, backgroundcolor=\color{white}][language=Python]
def hexagonal_num(n): 
	return n*(2*n - 1) \end{lstlisting}}   &
{\lstset{linewidth=3.5cm}
\begin{lstlisting}[frame=none, numbers=none, backgroundcolor=\color{white}][language=Python]


def hexagonal(n): 
  return n * (2*n - 1)\end{lstlisting}}   \\ \midrule 
{\lstset{linewidth=5.5cm}
\begin{lstlisting}[frame=none, numbers=none, backgroundcolor=\color{white}][language=Python]
Write a function to remove all elements from a given list present in another list.\end{lstlisting}}   &
{\lstset{linewidth=5.5cm}
\begin{lstlisting}[frame=none, numbers=none, backgroundcolor=\color{white}][language=Python]
def remove_elements(list1, list2):
    result = [x for x in list1 if x not in list2]
    return result\end{lstlisting}}   &
{\lstset{linewidth=3.5cm}
\begin{lstlisting}[frame=none, numbers=none, backgroundcolor=\color{white}][language=Python]
def intersection(lst1, lst2): 
    lst3 = [value for value in lst1 if value not in lst2] 
    return l\end{lstlisting}}   \\ \midrule 
{\lstset{linewidth=5.5cm}
\begin{lstlisting}[frame=none, numbers=none, backgroundcolor=\color{white}][language=Python]
Write a function to calculate the sum of the positive integers of n+(n-2)+(n-4)... (until n-x =< 0).\end{lstlisting}}   &
{\lstset{linewidth=5.5cm}
\begin{lstlisting}[frame=none, numbers=none, backgroundcolor=\color{white}][language=Python]
def sum_series(n):
  if n < 1:
    return 0
  else:
    return n + sum_series(n - 2)\end{lstlisting}}   &
{\lstset{linewidth=3.5cm}
\begin{lstlisting}[frame=none, numbers=none, backgroundcolor=\color{white}][language=Python]
def sum_series(n):
  if n < 1:
    return 0
  else:
    return n + sum_series(n - 2)\end{lstlisting}}   \\ \bottomrule
\end{tabular}
\caption{All questions flagged as being seen by models trained on the \stack 10 or more times within the MBPP benchmark (Part 2)}
\end{table*}

\endgroup
\begingroup

\setlength{\tabcolsep}{6pt} % Default value: 6pt
\renewcommand{\arraystretch}{0} % Default value: 1

\begin{table*}[]
\small
\begin{tabular}{L{4cm}|L{4cm}|L{3cm}}
\toprule
\textbf{Natural Language Question} & \textbf{Gold Program}                                                              & \textbf{Found 100\% Matches}   \\ \midrule 
{\lstset{linewidth=5.5cm}
\begin{lstlisting}[frame=none, numbers=none, backgroundcolor=\color{white}][language=Python]
from typing import List def filter_by_prefix(strings: List[str], prefix: str) -> List[str]: """ Filter an input list of strings only for ones that start with a given prefix. >>> filter_by_prefix([], 'a') [] >>> filter_by_prefix(['abc', 'bcd', 'cde', 'array'], 'a') ['abc', 'array'] """\end{lstlisting}}  &                                   
{\lstset{linewidth=5.5cm}
\begin{lstlisting}[frame=none, numbers=none, backgroundcolor=\color{white}][language=Python]
    return [x for x in strings if substring in x]\end{lstlisting}}   &
{\lstset{linewidth=3.5cm}
\begin{lstlisting}[frame=none, numbers=none, backgroundcolor=\color{white}][language=Python]
    return [tag for tag in tags if substr in tag]\end{lstlisting}}   \\ \midrule 
{\lstset{linewidth=5.5cm}
\begin{lstlisting}[frame=none, numbers=none, backgroundcolor=\color{white}][language=Python]
def greatest_common_divisor(a: int, b: int) -> int: """ Return a greatest common divisor of two integers a and b >>> greatest_common_divisor(3, 5) 1 >>> greatest_common_divisor(25, 15) 5 """\end{lstlisting}}   &
{\lstset{linewidth=5.5cm}
\begin{lstlisting}[frame=none, numbers=none, backgroundcolor=\color{white}][language=Python]
    while b:
        a, b = b, a % b
    return a\end{lstlisting}}   &
{\lstset{linewidth=3.5cm}
\begin{lstlisting}[frame=none, numbers=none, backgroundcolor=\color{white}][language=Python]
    while b:
        a, b = b, a % b
    return a\end{lstlisting}}   \\ \midrule 
{\lstset{linewidth=5.5cm}
\begin{lstlisting}[frame=none, numbers=none, backgroundcolor=\color{white}][language=Python]
from typing import List def concatenate(strings: List[str]) -> str: """ Concatenate list of strings into a single string >>> concatenate([]) '' >>> concatenate(['a', 'b', 'c']) 'abc' """\end{lstlisting}}   &
{\lstset{linewidth=5.5cm}
\begin{lstlisting}[frame=none, numbers=none, backgroundcolor=\color{white}][language=Python]
    return ''.join(strings)\end{lstlisting}}   &
{\lstset{linewidth=3.5cm}
\begin{lstlisting}[frame=none, numbers=none, backgroundcolor=\color{white}][language=Python]
    return ''.join(strings)\end{lstlisting}}   \\ \midrule 
{\lstset{linewidth=5.5cm}
\begin{lstlisting}[frame=none, numbers=none, backgroundcolor=\color{white}][language=Python]
from typing import List def filter_by_prefix(strings: List[str], prefix: str) -> List[str]: """ Filter an input list of strings only for ones that start with a given prefix. >>> filter_by_prefix([], 'a') [] >>> filter_by_prefix(['abc', 'bcd', 'cde', 'array'], 'a') ['abc', 'array'] """\end{lstlisting}}   &
{\lstset{linewidth=5.5cm}
\begin{lstlisting}[frame=none, numbers=none, backgroundcolor=\color{white}][language=Python]
    return [x for x in strings if x.startswith(prefix)]\end{lstlisting}}   &
{\lstset{linewidth=3.5cm}
\begin{lstlisting}[frame=none, numbers=none, backgroundcolor=\color{white}][language=Python]
    return [i for i in tests if i.startswith(prefix)]\end{lstlisting}}   \\ \midrule 
{\lstset{linewidth=5.5cm}
\begin{lstlisting}[frame=none, numbers=none, backgroundcolor=\color{white}][language=Python]
def encode_cyclic(s: str): """ returns encoded string by cycling groups of three characters. """ # split string to groups. Each of length 3. groups = [s[(3 * i):min((3 * i + 3), len(s))] for i in range((len(s) + 2) // 3)] # cycle elements in each group. Unless group has fewer elements than 3. groups = [(group[1:] + group[0]) if len(group) == 3 else group for group in groups] return "".join(groups) def decode_cyclic(s: str): """ takes as input string encoded with encode_cyclic function. Returns decoded string. """\end{lstlisting}}   &
{\lstset{linewidth=5.5cm}
\begin{lstlisting}[frame=none, numbers=none, backgroundcolor=\color{white}][language=Python]
    return encode_cyclic(encode_cyclic(s))\end{lstlisting}}   &
{\lstset{linewidth=3.5cm}
\begin{lstlisting}[frame=none, numbers=none, backgroundcolor=\color{white}][language=Python]
    return encodedValue(encode(value));\end{lstlisting}}   \\ \midrule 
{\lstset{linewidth=5.5cm}
\begin{lstlisting}[frame=none, numbers=none, backgroundcolor=\color{white}][language=Python]
def add(x: int, y: int): """Add two numbers x and y >>> add(2, 3) 5 >>> add(5, 7) 12 """\end{lstlisting}}   &
{\lstset{linewidth=5.5cm}
\begin{lstlisting}[frame=none, numbers=none, backgroundcolor=\color{white}][language=Python]
    return x + y\end{lstlisting}}   &
{\lstset{linewidth=3.5cm}
\begin{lstlisting}[frame=none, numbers=none, backgroundcolor=\color{white}][language=Python]
    return x + y\end{lstlisting}}   \\ \bottomrule
\end{tabular}
\caption{All questions flagged as being seen by models trained on the \pile 10 or more times within the HumanEval benchmark (Part 1)}
\end{table*}

\endgroup
\begingroup

\setlength{\tabcolsep}{6pt} % Default value: 6pt
\renewcommand{\arraystretch}{0} % Default value: 1

\begin{table*}[]
\small
\begin{tabular}{L{4cm}|L{4cm}|L{3cm}}
\toprule
\textbf{Natural Language Question} & \textbf{Gold Program}                                                              & \textbf{Found 100\% Matches}                                                       \\ \midrule
{\lstset{linewidth=5.5cm}
\begin{lstlisting}[frame=none, numbers=none, backgroundcolor=\color{white}][language=Python]
def same_chars(s0: str, s1: str): """ Check if two words have the same characters. >>> same_chars('eabcdzzzz', 'dddzzzzzzzddeddabc') True >>> same_chars('abcd', 'dddddddabc') True >>> same_chars('dddddddabc', 'abcd') True >>> same_chars('eabcd', 'dddddddabc') False >>> same_chars('abcd', 'dddddddabce') False >>> same_chars('eabcdzzzz', 'dddzzzzzzzddddabc') False """\end{lstlisting}}   &
{\lstset{linewidth=5.5cm}
\begin{lstlisting}[frame=none, numbers=none, backgroundcolor=\color{white}][language=Python]
    return set(s0) == set(s1)\end{lstlisting}}   &
{\lstset{linewidth=3.5cm}
\begin{lstlisting}[frame=none, numbers=none, backgroundcolor=\color{white}][language=Python]
    return set(l1) == set(l2)\end{lstlisting}}   \\ \midrule 
{\lstset{linewidth=5.5cm}
\begin{lstlisting}[frame=none, numbers=none, backgroundcolor=\color{white}][language=Python]
def multiply(a, b): """Complete the function that takes two integers and returns the product of their unit digits. Assume the input is always valid. Examples: multiply(148, 412) should return 16. multiply(19, 28) should return 72. multiply(2020, 1851) should return 0. multiply(14,-15) should return 20. """\end{lstlisting}}   &
{\lstset{linewidth=5.5cm}
\begin{lstlisting}[frame=none, numbers=none, backgroundcolor=\color{white}][language=Python]
    return abs(a % 10) * abs(b % 10)\end{lstlisting}}   &
{\lstset{linewidth=3.5cm}
\begin{lstlisting}[frame=none, numbers=none, backgroundcolor=\color{white}][language=Python]
   return abs(fa - f0) < abs(fb - f0)\end{lstlisting}}   \\ \bottomrule
\end{tabular}
\caption{All questions flagged as being seen by models trained on the \pile 10 or more times within the HumanEval benchmark (Part 2)}
\end{table*}

\endgroup
\begingroup

\setlength{\tabcolsep}{6pt} % Default value: 6pt
\renewcommand{\arraystretch}{0} % Default value: 

\begin{table*}[]
\small
\begin{tabular}{L{4cm}|L{4cm}|L{3cm}}
\toprule
\textbf{Natural Language Question} & \textbf{Gold Program}                                                              & \textbf{Found 100\% Matches}                                                       \\ \midrule
{\lstset{linewidth=5.5cm}
\begin{lstlisting}[frame=none, numbers=none, backgroundcolor=\color{white}, backgroundcolor=\color{white}][language=Python]
def strlen(string: str) -> int: """ Return length of given string >>> strlen('') 0 >>> strlen('abc') 3 """\end{lstlisting}}   &
{\lstset{linewidth=5.5cm}
\begin{lstlisting}[frame=none, numbers=none, backgroundcolor=\color{white}, backgroundcolor=\color{white}][language=Python]
    return len(string)\end{lstlisting}}   &
{\lstset{linewidth=3.5cm}
\begin{lstlisting}[frame=none, numbers=none, backgroundcolor=\color{white}][language=Python]
    return len(string)\end{lstlisting}}   \\ \midrule 
{\lstset{linewidth=5.5cm}
\begin{lstlisting}[frame=none, numbers=none, backgroundcolor=\color{white}][language=Python]
def flip_case(string: str) -> str: """ For a given string, flip lowercase characters to uppercase and uppercase to lowercase. >>> flip_case('Hello') 'hELLO' """\end{lstlisting}}   &
{\lstset{linewidth=5.5cm}
\begin{lstlisting}[frame=none, numbers=none, backgroundcolor=\color{white}][language=Python]
    return string.swapcase()\end{lstlisting}}   &
{\lstset{linewidth=3.5cm}
\begin{lstlisting}[frame=none, numbers=none, backgroundcolor=\color{white}][language=Python]
    return string_.swapcase()\end{lstlisting}}   \\ \midrule  
{\lstset{linewidth=5.5cm}
\begin{lstlisting}[frame=none, numbers=none, backgroundcolor=\color{white}][language=Python]
def sort_third(l: list): """This function takes a list l and returns a list l' such that l' is identical to l in the indicies that are not divisible by three, while its values at the indicies that are divisible by three are equal to the values of the corresponding indicies of l, but sorted. >>> sort_third([1, 2, 3]) [1, 2, 3] >>> sort_third([5, 6, 3, 4, 8, 9, 2]) [2, 6, 3, 4, 8, 9, 5] """\end{lstlisting}}   & 
{\lstset{linewidth=5.5cm}
\begin{lstlisting}[frame=none, numbers=none, backgroundcolor=\color{white}][language=Python]
    l = list(l)
    l[::3] = sorted(l[::3])
    return l\end{lstlisting}}   &
{\lstset{linewidth=3.5cm}
\begin{lstlisting}[frame=none, numbers=none, backgroundcolor=\color{white}][language=Python]
    l = list(str(n))
    s = sorted(l)
    return s != l\end{lstlisting}}   \\ \midrule  
{\lstset{linewidth=5.5cm}
\begin{lstlisting}[frame=none, numbers=none, backgroundcolor=\color{white}][language=Python]
def fizz_buzz(n: int): """Return the number of times the digit 7 appears in integers less than n which are divisible by 11 or 13. >>> fizz_buzz(50) 0 >>> fizz_buzz(78) 2 >>> fizz_buzz(79) 3 """\end{lstlisting}}   & 
{\lstset{linewidth=5.5cm}
\begin{lstlisting}[frame=none, numbers=none, backgroundcolor=\color{white}][language=Python]
    ns = []
    for i in range(n):
        if i % 11 == 0 or i % 13 == 0:
            ns.append(i)
    s = ''.join(list(map(str, ns)))
    ans = 0
    for c in s:
        ans += (c == '7')
    return ans\end{lstlisting}}   &
{\lstset{linewidth=3.5cm}
\begin{lstlisting}[frame=none, numbers=none, backgroundcolor=\color{white}][language=Python]
    fizzy = []
    for i in range(1, n+1):
        if i % 3 == 0 and i % 5 == 0:
            fizzy.append('FizzBuzz')
        elif i % 3 == 0:
            fizzy.append('Fizz')
        elif i % 5 == 0:\end{lstlisting}}   \\ \midrule  
{\lstset{linewidth=5.5cm}
\begin{lstlisting}[frame=none, numbers=none, backgroundcolor=\color{white}][language=Python]
	
from typing import List, Optional def longest(strings: List[str]) -> Optional[str]: """ Out of list of strings, return the longest one. Return the first one in case of multiple strings of the same length. Return None in case the input list is empty. >>> longest([]) >>> longest(['a', 'b', 'c']) 'a' >>> longest(['a', 'bb', 'ccc']) 'ccc' """\end{lstlisting}}   & 
{\lstset{linewidth=5.5cm}
\begin{lstlisting}[frame=none, numbers=none, backgroundcolor=\color{white}][language=Python]
    if not strings:
        return None

    maxlen = max(len(x) for x in strings)
    for s in strings:
        if len(s) == maxlen:
            return s\end{lstlisting}}   &
{\lstset{linewidth=3.5cm}
\begin{lstlisting}[frame=none, numbers=none, backgroundcolor=\color{white}][language=Python]
    if not strings:
        return ''
    prefix = strings[0]
    for s in strings:
        if len(s) < len(prefix):
            prefix = prefix[:len(s\end{lstlisting}}   \\ \midrule  
{\lstset{linewidth=5.5cm}
\begin{lstlisting}[frame=none, numbers=none, backgroundcolor=\color{white}][language=Python]
def how_many_times(string: str, substring: str) -> int: """ Find how many times a given substring can be found in the original string. Count overlaping cases. >>> how_many_times('', 'a') 0 >>> how_many_times('aaa', 'a') 3 >>> how_many_times('aaaa', 'aa') 3 """\end{lstlisting}}   & 
{\lstset{linewidth=5.5cm}
\begin{lstlisting}[frame=none, numbers=none, backgroundcolor=\color{white}][language=Python]
    times = 0

    for i in range(len(string) - len(substring) + 1):
        if string[i:i+len(substring)] == substring:
            times += 1

    return times\end{lstlisting}}   &
{\lstset{linewidth=3.5cm}
\begin{lstlisting}[frame=none, numbers=none, backgroundcolor=\color{white}][language=Python]
    result = 0
    for i in range(len(string) - len(sub_string) + 1):
        if string[i:i+len(sub_string)] == sub_string:
            result += 1
    return res\end{lstlisting}}\\ \bottomrule 

\end{tabular}
\caption{All questions flagged as being seen by models trained on the \stack 10 or more times within the HumanEval benchmark (Part 1)}
\label{tab:seen-end}
\end{table*}

\endgroup
\begingroup

\setlength{\tabcolsep}{6pt} % Default value: 6pt
\renewcommand{\arraystretch}{0} % Default value: 1

\begin{table*}[]
\small
\begin{tabular}{L{4cm}|L{4cm}|L{3cm}}
\toprule
\textbf{Natural Language Question} & \textbf{Gold Program}                                                              & \textbf{Found 100\% Matches}   \\ \midrule  
{\lstset{linewidth=5.5cm}
\begin{lstlisting}[frame=none, numbers=none, backgroundcolor=\color{white}][language=Python]
	
from typing import List def sort_numbers(numbers: str) -> str: """ Input is a space-delimited string of numberals from 'zero' to 'nine'. Valid choices are 'zero', 'one', 'two', 'three', 'four', 'five', 'six', 'seven', 'eight' and 'nine'. Return the string with numbers sorted from smallest to largest >>> sort_numbers('three one five') 'one three five' """\end{lstlisting}}   & 
{\lstset{linewidth=5.5cm}
\begin{lstlisting}[frame=none, numbers=none, backgroundcolor=\color{white}][language=Python]
    value_map = {
        'zero': 0,
        'one': 1,
        'two': 2,
        'three': 3,
        'four': 4,
        'five': 5,
        'six': 6,
        'seven': 7,
        'eight': 8,
        'nine': 9
    }
    return ' '.join(sorted([x for x in numbers.split(' ') if x], key=lambda x: value_map[x]))\end{lstlisting}}   &
{\lstset{linewidth=3.5cm}
\begin{lstlisting}[frame=none, numbers=none, backgroundcolor=\color{white}][language=Python]
nine
    number_map = {
        'zero': 0,
        'one': 1,
        'two': 2,
        'three': 3,
        'four': 4,
        'five': 5,
        'six': 6,
        'seven': 7,
        'eight': 8,
        'nine': 9
    }

    # Override the init method and use the kwarg "name" to set a string that will be sl\end{lstlisting}}   \\ \midrule 
{\lstset{linewidth=5.5cm}
\begin{lstlisting}[frame=none, numbers=none, backgroundcolor=\color{white}][language=Python]
	
def tri(n): """Everyone knows Fibonacci sequence, it was studied deeply by mathematicians in the last couple centuries. However, what people don't know is Tribonacci sequence. Tribonacci sequence is defined by the recurrence: tri(1) = 3 tri(n) = 1 + n / 2, if n is even. tri(n) = tri(n - 1) + tri(n - 2) + tri(n + 1), if n is odd. For example: tri(2) = 1 + (2 / 2) = 2 tri(4) = 3 tri(3) = tri(2) + tri(1) + tri(4) = 2 + 3 + 3 = 8 You are given a non-negative integer number n, you have to a return a list of the first n + 1 numbers of the Tribonacci sequence. Examples: tri(3) = [1, 3, 2, 8] """\end{lstlisting}}   &
{\lstset{linewidth=5.5cm}
\begin{lstlisting}[frame=none, numbers=none, backgroundcolor=\color{white}][language=Python]
    if n == 0:
        return [1]
    my_tri = [1, 3]
    for i in range(2, n + 1):
        if i % 2 == 0:
            my_tri.append(i / 2 + 1)
        else:
            my_tri.append(my_tri[i - 1] + my_tri[i - 2] + (i + 3) / 2)
    return my_tri\end{lstlisting}}   &
{\lstset{linewidth=3.5cm}
\begin{lstlisting}[frame=none, numbers=none, backgroundcolor=\color{white}][language=Python]
= len(my_list)
    if list_len == 0:
        return None

    bool_list = []
    for i in range(list_len):
        if my_list[i] % 2 == 0:
            bool_list.append(True)
        else:
            bool_list.append(False)
    return (bool_list)\end{lstlisting}}   \\ \midrule  
{\lstset{linewidth=5.5cm}
\begin{lstlisting}[frame=none, numbers=none, backgroundcolor=\color{white}][language=Python]
def check_if_last_char_is_a_letter(txt): ''' Create a function that returns True if the last character of a given string is an alphabetical character and is not a part of a word, and False otherwise. Note: "word" is a group of characters separated by space. Examples: check_if_last_char_is_a_letter("apple pie") -> False check_if_last_char_is_a_letter("apple pi e") -> True check_if_last_char_is_a_letter("apple pi e ") -> False check_if_last_char_is_a_letter("") -> False '''\end{lstlisting}}   & 
{\lstset{linewidth=5.5cm}
\begin{lstlisting}[frame=none, numbers=none, backgroundcolor=\color{white}][language=Python]
    check = txt.split(' ')[-1]
    return True if len(check) == 1 and (97 <= ord(check.lower()) <= 122) else False\end{lstlisting}}   &
{\lstset{linewidth=3.5cm}
\begin{lstlisting}[frame=none, numbers=none, backgroundcolor=\color{white}][language=Python]
     prefix = re.split(r'[\.\_]', id)[0]
        return True if len(prefix) == 6 and int(prefix) > 101000 else False\end{lstlisting}}   \\ \midrule  
{\lstset{linewidth=5.5cm}
\begin{lstlisting}[frame=none, numbers=none, backgroundcolor=\color{white}][language=Python]
	
def can_arrange(arr): """Create a function which returns the largest index of an element which is not greater than or equal to the element immediately preceding it. If no such element exists then return -1. The given array will not contain duplicate values. Examples: can_arrange([1,2,4,3,5]) = 3 can_arrange([1,2,3]) = -1 """\end{lstlisting}}   & 
{\lstset{linewidth=5.5cm}
\begin{lstlisting}[frame=none, numbers=none, backgroundcolor=\color{white}][language=Python]
    ind=-1
    i=1
    while i<len(arr):
      if arr[i]<arr[i-1]:
        ind=i
      i+=1
    return ind\end{lstlisting}}   &
{\lstset{linewidth=3.5cm}
\begin{lstlisting}[frame=none, numbers=none, backgroundcolor=\color{white}][language=Python]
    i = 1
    while i < len(arr):
        if arr[i-1] < arr[i]:
            i += 1
        elif arr[i\end{lstlisting}}   \\ \bottomrule

\end{tabular}
\caption{All questions flagged as being seen by models trained on the \stack 10 or more times within the HumanEval benchmark (Part 2)}
\end{table*}

\endgroup
\begingroup

\setlength{\tabcolsep}{6pt} % Default value: 6pt
\renewcommand{\arraystretch}{0} % Default value: 1

\begin{table*}[]
\small
\begin{tabular}{L{4cm}|L{4cm}|L{3cm}}
\toprule
\textbf{Natural Language Question} & \textbf{Gold Program}                                                              & \textbf{Found 100\% Matches}                                                       \\ \midrule 
{\lstset{linewidth=5.5cm}
\begin{lstlisting}[frame=none, numbers=none, backgroundcolor=\color{white}][language=Python]
def is_equal_to_sum_even(n): """Evaluate whether the given number n can be written as the sum of exactly 4 positive even numbers Example is_equal_to_sum_even(4) == False is_equal_to_sum_even(6) == False is_equal_to_sum_even(8) == True """\end{lstlisting}}   &
{\lstset{linewidth=5.5cm}
\begin{lstlisting}[frame=none, numbers=none, backgroundcolor=\color{white}][language=Python]
    return n%2 == 0 and n >= 8\end{lstlisting}}   &
{\lstset{linewidth=3.5cm}
\begin{lstlisting}[frame=none, numbers=none, backgroundcolor=\color{white}][language=Python]
5
    return n2 == 7 and n1 >= \end{lstlisting}}   \\ \midrule 
{\lstset{linewidth=5.5cm}
\begin{lstlisting}[frame=none, numbers=none, backgroundcolor=\color{white}][language=Python]
def car_race_collision(n: int): """ Imagine a road that's a perfectly straight infinitely long line. n cars are driving left to right; simultaneously, a different set of n cars are driving right to left. The two sets of cars start out being very far from each other. All cars move in the same speed. Two cars are said to collide when a car that's moving left to right hits a car that's moving right to left. However, the cars are infinitely sturdy and strong; as a result, they continue moving in their trajectory as if they did not collide. This function outputs the number of such collisions. """\end{lstlisting}}   &
{\lstset{linewidth=5.5cm}
\begin{lstlisting}[frame=none, numbers=none, backgroundcolor=\color{white}][language=Python]
    return n**2\end{lstlisting}}   &
{\lstset{linewidth=3.5cm}
\begin{lstlisting}[frame=none, numbers=none, backgroundcolor=\color{white}][language=Python]
    return n**2\end{lstlisting}}   \\ \midrule
{\lstset{linewidth=5.5cm}
\begin{lstlisting}[frame=none, numbers=none, backgroundcolor=\color{white}][language=Python]
def reverse_delete(s,c): """Task We are given two strings s and c, you have to deleted all the characters in s that are equal to any character in c then check if the result string is palindrome. A string is called palindrome if it reads the same backward as forward. You should return a tuple containing the result string and True/False for the check. Example For s = "abcde", c = "ae", the result should be ('bcd',False) For s = "abcdef", c = "b" the result should be ('acdef',False) For s = "abcdedcba", c = "ab", the result should be ('cdedc',True) """\end{lstlisting}}    &
{\lstset{linewidth=5.5cm}
\begin{lstlisting}[frame=none, numbers=none, backgroundcolor=\color{white}][language=Python]
    s = ''.join([char for char in s if char not in c])
    return (s,s[::-1] == s)\end{lstlisting}}   &
{\lstset{linewidth=3.5cm}
\begin{lstlisting}[frame=none, numbers=none, backgroundcolor=\color{white}][language=Python]
)
    s = ''.join([char for char in s if char not in dashes])
    return s.lower()\end{lstlisting}}   \\ \midrule 
{\lstset{linewidth=5.5cm}
\begin{lstlisting}[frame=none, numbers=none, backgroundcolor=\color{white}][language=Python]
def minSubArraySum(nums): """ Given an array of integers nums, find the minimum sum of any non-empty sub-array of nums. Example minSubArraySum([2, 3, 4, 1, 2, 4]) == 1 minSubArraySum([-1, -2, -3]) == -6 """\end{lstlisting}}   &
{\lstset{linewidth=5.5cm}
\begin{lstlisting}[frame=none, numbers=none, backgroundcolor=\color{white}][language=Python]
    max_sum = 0
    s = 0
    for num in nums:
        s += -num
        if (s < 0):
            s = 0
        max_sum = max(s, max_sum)
    if max_sum == 0:
        max_sum = max(-i for i in nums)
    min_sum = -max_sum
    return min_sum\end{lstlisting}}   &
{\lstset{linewidth=3.5cm}
\begin{lstlisting}[frame=none, numbers=none, backgroundcolor=\color{white}][language=Python]
        max_sum, sub_sum = nums[0], nums[0]
        for num in nums[1:]:
            s = num
            if sub_sum > 0:
                s += sub_sum
            max_sum = max(s, max_sum)
            sub_sum = s
        return max_sum\end{lstlisting}}   \\ \bottomrule

\end{tabular}
\caption{All questions flagged as being seen by models trained on the \stack 10 or more times within the HumanEval benchmark (Part 3)}
\end{table*}

\endgroup
\begingroup

\setlength{\tabcolsep}{6pt} % Default value: 6pt
\renewcommand{\arraystretch}{0} % Default value: 1

\begin{table*}[]
\small
\begin{tabular}{L{4cm}|L{4cm}|L{3cm}}
\toprule
\textbf{Natural Language Question} & \textbf{Gold Program}                                                              & \textbf{Found 100\% Matches}                                                       \\ \midrule 
{\lstset{linewidth=5.5cm}
\begin{lstlisting}[frame=none, numbers=none, backgroundcolor=\color{white}][language=Python]
def max_fill(grid, capacity): import math """ You are given a rectangular grid of wells. Each row represents a single well, and each 1 in a row represents a single unit of water. Each well has a corresponding bucket that can be used to extract water from it, and all buckets have the same capacity. Your task is to use the buckets to empty the wells. Output the number of times you need to lower the buckets. Example 1: Input: grid : [[0,0,1,0], [0,1,0,0], [1,1,1,1]] bucket_capacity : 1 Output: 6 Example 2: Input: grid : [[0,0,1,1], [0,0,0,0], [1,1,1,1], [0,1,1,1]] bucket_capacity : 2 Output: 5 Example 3: Input: grid : [[0,0,0], [0,0,0]] bucket_capacity : 5 Output: 0 Constraints: * all wells have the same length * 1 <= grid.length <= 10^2 * 1 <= grid[:,1].length <= 10^2 * grid[i][j] -> 0 | 1 * 1 <= capacity <= 10 """\end{lstlisting}}   &
{\lstset{linewidth=5.5cm}
\begin{lstlisting}[frame=none, numbers=none, backgroundcolor=\color{white}][language=Python]
    return sum([math.ceil(sum(arr)/capacity) for arr in grid])\end{lstlisting}}   &
{\lstset{linewidth=3.5cm}
\begin{lstlisting}[frame=none, numbers=none, backgroundcolor=\color{white}][language=Python]
        return sum(int(math.ceil(ntasks)) for ntasks in rows)\end{lstlisting}}   \\ \midrule
{\lstset{linewidth=5.5cm}
\begin{lstlisting}[frame=none, numbers=none, backgroundcolor=\color{white}][language=Python]
    def add(x: int, y: int): """Add two numbers x and y >>> add(2, 3) 5 >>> add(5, 7) 12 """\end{lstlisting}}    &
{\lstset{linewidth=5.5cm}
\begin{lstlisting}[frame=none, numbers=none, backgroundcolor=\color{white}][language=Python]
    return x + y\end{lstlisting}}   &
{\lstset{linewidth=3.5cm}
\begin{lstlisting}[frame=none, numbers=none, backgroundcolor=\color{white}][language=Python]
    return x + y\end{lstlisting}}   \\ \midrule 
{\lstset{linewidth=5.5cm}
\begin{lstlisting}[frame=none, numbers=none, backgroundcolor=\color{white}][language=Python]
def do_algebra(operator, operand): """ Given two lists operator, and operand. The first list has basic algebra operations, and the second list is a list of integers. Use the two given lists to build the algebric expression and return the evaluation of this expression. The basic algebra operations: Addition ( + ) Subtraction ( - ) Multiplication ( * ) Floor division ( // ) Exponentiation ( ** ) Example: operator['+', '*', '-'] array = [2, 3, 4, 5] result = 2 + 3 * 4 - 5 => result = 9 Note: The length of operator list is equal to the length of operand list minus one. Operand is a list of of non-negative integers. Operator list has at least one operator, and operand list has at least two operands. """\end{lstlisting}}   &
{\lstset{linewidth=5.5cm}
\begin{lstlisting}[frame=none, numbers=none, backgroundcolor=\color{white}][language=Python]
    expression = str(operand[0])
    for oprt, oprn in zip(operator, operand[1:]):
        expression+= oprt + str(oprn)
    return eval(expression)\end{lstlisting}}   &
{\lstset{linewidth=3.5cm}
\begin{lstlisting}[frame=none, numbers=none, backgroundcolor=\color{white}][language=Python]
    result = str(self.operand[0])
    for operator, operand in zip(self.operator, self.operand[1:]):
      result += ' {} {}'.format(operator, operan\end{lstlisting}}\\ \midrule 
{\lstset{linewidth=5.5cm}
\begin{lstlisting}[frame=none, numbers=none, backgroundcolor=\color{white}][language=Python]
def encode(message): """ Write a function that takes a message, and encodes in such a way that it swaps case of all letters, replaces all vowels in the message with the letter that appears 2 places ahead of that vowel in the english alphabet. Assume only letters. Examples: >>> encode('test') 'TGST' >>> encode('This is a message') 'tHKS KS C MGSSCGG' """\end{lstlisting}}   &
{\lstset{linewidth=5.5cm}
\begin{lstlisting}[frame=none, numbers=none, backgroundcolor=\color{white}][language=Python]
    vowels = "aeiouAEIOU"
    vowels_replace = dict([(i, chr(ord(i) + 2)) for i in vowels])
    message = message.swapcase()
    return ''.join([vowels_replace[i] if i in vowels else i for i in message])\end{lstlisting}}   &
{\lstset{linewidth=3.5cm}
\begin{lstlisting}[frame=none, numbers=none, backgroundcolor=\color{white}][language=Python]
        vovels = "aeiouAEIOU"
        stack = []
        for ch in s:
            if ch in vovels:
                stack.append(ch)
        result = []
        for i in s:
            if i in vovels\end{lstlisting}}   \\ \bottomrule

\end{tabular}
\caption{All questions flagged as being seen by models trained on the \stack 10 or more times within the HumanEval benchmark (Part 4)}
\label{tab:seen-end}
\end{table*}

\endgroup

\end{document}